\documentclass[12pt]{article}

\usepackage{amsmath,amssymb,amsfonts,amscd,mathrsfs,mathtools,bbold, braket,bm,xfrac}
\usepackage{xcolor}
\definecolor{darkblue}{rgb}{0.1,0.1,.7}
\usepackage[colorlinks, linkcolor=darkblue, citecolor=darkblue, urlcolor=darkblue, linktocpage]{hyperref} 
\usepackage{geometry}
\usepackage{tablefootnote,colortbl}

\usepackage{braket}

\usepackage{cancel}
\usepackage{arcs}

\usepackage{arydshln}
\usepackage{booktabs,multirow}
\usepackage{amssymb}
\usepackage{fancyvrb}

\usepackage{tikz-feynman}
 \tikzfeynmanset{compat=1.0.0}

 \usepackage[square, comma, compress,numbers]{natbib}
\usepackage{subcaption}

\definecolor{myorange}{RGB}{199,146,32}

\definecolor{Gray1}{gray}{0.97}
\definecolor{Gray2}{gray}{0.9}
\definecolor{LightCyan}{rgb}{0.88,1,1}
\definecolor{blu}{rgb}{0,0,1}

\usepackage{ragged2e}
\usepackage{array}
\newcolumntype{L}[1]{>{\raggedright\let\newline\\\arraybackslash\hspace{0pt}}m{#1}}
\newcolumntype{C}[1]{>{\centering\let\newline\\\arraybackslash\hspace{0pt}}m{#1}}
\newcolumntype{R}[1]{>{\raggedleft\let\newline\\\arraybackslash\hspace{0pt}}m{#1}}

\usepackage{dsfont} 
\usepackage{tcolorbox}
\usepackage{booktabs}

\usepackage{titlesec}
\titleformat*{\section}{\large\bfseries}
\titleformat*{\subsection}{\normalsize\bfseries}
\titleformat*{\subsubsection}{\normalsize\it}
\titleformat*{\paragraph}{\normalsize\bfseries}
\titleformat*{\subparagraph}{\normalsize\bfseries}

\def\D {{\Delta_T}}

\newcommand{\reef}[1]{(\ref{#1})}

\def\eps{\epsilon}

\newcommand{\beq}{\begin{equation}} 
\newcommand{\eeq}{\end{equation}}

\def\bZ {\mathbb{Z}}

\def\calO {{\cal O}}

\def\bZ {\mathbb{Z}}

\def\ge{\geqslant}
\def\le{\leqslant}
\def\geq{\geqslant}
\def\leq{\leqslant}
\newcommand{\diffop}[2]{\ifthenelse{\equal{#2}{1}}{\frac{\mrm{d}}{\mrm{d} #1}}{\frac{\mrm{d}^#2}{\mrm{d} #1^#2}}}

\newcommand{\overbar}[1]{\mkern 1.5mu\overline{\mkern-1.5mu#1\mkern-1.5mu}\mkern 1.5mu} 

\newcommand\myline{
	\,\tikz[baseline]\draw[ thick,dashed](0,-\dp\strutbox)--(0,\ht\strutbox);\,
}

\newcommand{\mrm}[1]{{\mathrm #1}}

\usepackage[normalem]{ulem}

\newcommand{\be}{\begin{equation}}
\newcommand{\ee}{\end{equation}}
\def\bea#1\eea{\begin{align}#1\end{align}}

  \def\th{\theta}

\newcommand{\matrixel}[3]{\left< #1 \vphantom{#2#3} \right|
	#2 \left| #3 \vphantom{#1#2} \right>} 

\usepackage{accents}
\newlength{\dhatheight}

\numberwithin{equation}{section}
\usepackage{tocloft}
\begin{document}

\vspace*{-.6in} \thispagestyle{empty}
\begin{flushright}
IPPP/24/79
\end{flushright}
\vspace{1cm} {\Large
\begin{center}
\textbf{
Testing the RG-flow $M(3,10)+\phi_{1,7}\to M(3,8)$ \\[.2cm] with Hamiltonian Truncation
 }
\end{center}}
\vspace{1cm}
\begin{center}

{\bf  Olivier Delouche$^{a,b}$,  Joan Elias~Mir\'o$^{b}$, James Ingoldby$^{c}$ } \\[1cm] 
 {$^a$  Université de Genève,  24 quai Ernest-Ansermet, 1211 Genève 4, Switzerland  \\ 
  $^b$ The Abdus Salam ICTP,    Strada Costiera 11, 34151, Trieste, Italy  \\ 
   $^c$  Institute for Particle Physics Phenomenology, Durham University, Durham DH1 3LE, UK  \\ 
 }
\vspace{1cm}

\abstract{ 

\bigskip
\noindent 

Hamiltonian Truncation (HT) methods provide a powerful numerical approach for investigating strongly coupled QFTs.
In this work, we develop HT techniques to analyse  a specific  Renormalization Group (RG) flow recently proposed in 
Refs.~\cite{Fei:2014xta,Klebanov:2022syt,Katsevich:2024jgq}.
These studies put forward Ginzburg-Landau descriptions for the conformal minimal models $M(3,10)$ and $M(3,8)$, as well as the RG-flow connecting them. Specifically, the RG-flow is defined by deforming the  $M(3,10)$ with the relevant primary operator  $\phi_{1,7}$ (whose indices denote its position in the Kac table), yielding $M(3,10)+ \phi_{1,7}$.
From the perspective of HT, realising  such an  RG-flow presents significant challenges, as the $\phi_{1,7}$ deformation requires renormalizing the UV theory up to third order in the coupling constant of the deformation. 
In this study, we carry out the necessary calculations to formulate HT  for this theory  and  numerically investigate the spectrum of $M(3,10)+ \phi_{1,7}$ in the  large coupling regime, finding strong evidence in favor of the proposed flow. 

}

\vspace{3cm}
\end{center}

 \vfill
 {
  \flushright
 \today 
}

\newpage 


 
 
\tableofcontents

\section{Introduction}
\label{intro}

Besides their success in describing critical phenomena, Conformal Minimal Models in two spacetime dimensions~\cite{Belavin:1984vu}, and the Renormalization Group (RG) flows connecting them, serve as valuable examples for understanding the non-perturbative regime of Quantum Field Theory (QFT).  In this work, we investigate RG-flows that start from a particular Minimal Model nonperturbatively, using Hamiltonian Truncation.

Minimal Models are denoted by $M(p,q)$ where $p$ and $q$ are  co-prime integers, i.e. without common divisors. 
The central charge of these CFTs is given by $c(p,q)=1-6(p-q)^2/(pq)$, and there are $(p-1)(q-1)/2$ distinct Virasoro primary operators $\phi_{r,s}$ with holomorphic dimensions
$
h_{r,s} =h_{q-r,p-s}=\frac{(rq -sp)^2-(p-q)^2}{4pq}
$
where $r=1,\dots, p-1$ and $  s=1,\dots,q-1$.
The  operator with lowest dimension has $h_\text{min}=\frac{1-(p-q)^2}{4pq}$.  

When  $q=p+1$, the Minimal Model is \emph{unitary} with $h_\text{min} =0$, corresponding to the identity operator. Two well-known examples of unitary Minimal Models are the two-dimensional Ising CFT $M(3,4)$, 
and the tricritical Ising CFT $M(4,5)$. Other choices of coprime indices such that $|p-q|> 1$ correspond to \emph{non-unitary} Minimal Models,
because these CFTs have a negative central charge and operators with negative scaling dimension. 
Nevertheless they have a number of physical applications. 
A prime example in this class is the two-dimensional Lee-Yang CFT $M(2,5)$, which describes the scaling behavior of the Yang-Lee edge singularity~\cite{Fisher:1978pf,Cardy:1985yy} -- see Ref.~\cite{Cardy:2023lha} for a recent review. 

The unitary Minimal Models admit a semi-classical or Ginzburg-Landau (GL) description -- see Ref.~\cite{Mussardo:2010mgq} for a textbook introduction.  
They can be described by the effective action
\be
S_{m+1,m+2} = \int d^dx \left( \frac{1}{2}(\partial_\mu \phi)^2 + \frac{g}{(2m)!}\phi^{2m} \right) \, . 
\ee
where $m=2,3,4 \dots$ and  the coefficients of the terms  $\phi^{2s}$ with $s<m$ are tuned to zero.
Because the scalar $\phi$ is dimensionless in $d=2$ spacetime dimensions, $g$ has dimensions of mass squared and 
therefore the theory is strongly coupled in the infrared (IR), i.e. for squared-energy scales below $g$. 
For $m=2$, the operator $\phi^4$ becomes classically marginal at $d=4$. 
It is then possible to consider the theory at $d=4-\eps$ dimensions, tune to the Wilson-Fisher~\cite{Wilson:1971dc} fixed-point, and perform perturbative calculations  in
$\epsilon$ of various anomalous dimensions of the operators in the theory. 
Extrapolating these calculations all the way to $d=2$ shows a nice agreement between the  $S_{3,4}$ predictions and the $M(3,4)$ minimal model. 
Evidence for this  interpretation is also found in numerical calculations performed directly at $d=2$ spacetime dimensions, such as: Lattice Monte Carlo~\cite{Schaich:2009jk}, Hamiltonian Truncation~\cite{Rychkov:2014eea,Elias-Miro:2017xxf,Lajer:2023unt} or Borel resummations~\cite{Serone:2018gjo}.

The Lee-Yang model $M(2,5)$, also admits a GL description \cite{Fisher:1978pf}.
It is described by the effective action
\be
S_{LY}= \int d^dx \left( \frac{1}{2}(\partial_\mu \phi)^2 + \frac{ig}{3!}\phi^{3} \right) \, .  \label{GLLY}
\ee
The coupling $ig$ is purely imaginary, rendering the theory non-unitary but without runaway unstable directions.~\footnote{The Hamiltonian of this theory is not Hermitian but  satisfies   $[X,H]=0$ for an antilinear symmetry operator $X$.  The last equation guarantees that the eigenvalues of $H$ are either real or appear in complex conjugate pairs. In our case $X$ can be chosen to be an anti-linear ${\cal PT}$-transformation. See the following sections and the review \cite{Bender:2018pbv}.}
The dimension of $\phi^3$ is $3(d-2)/2$. This operator is classically marginal at $d=6$ spacetime dimensions. Again, it is possible to perform Wilson-Fisher 
perturbative expansions in $\epsilon$ for $d=6-\eps$ to compute the anomalous dimensions of the first few lowest lying operators. Extrapolations of these results are in good agreement with the $M(2,5)$ predictions~\cite{Borinsky:2021jdb}.
The GL interpretation \reef{GLLY} is also   supported by Hamiltonian Truncation  calculations in $d=2$, by computing   the RG-flow from Ising to Lee-Yang~\cite{Fonseca:2001dc,Xu:2022mmw} or by directly diagonalising the Hamiltonian of the theory in  \reef{GLLY} and tuning the couplings to criticality~\cite{Lencses:2024wib}.~\footnote{Hamiltonian Truncation is often called TCSA or TFFSA in this context.}

While several generalizations of the GL description of Minimal Models exist~\cite{Fei:2014xta,Klebanov:2022syt,Katsevich:2024jgq,tesi-amoruso,Zambelli:2016cbw},
a general   classification is lacking. 
In this work we deal with the particular generalization  introduced in Refs.~\cite{Fei:2014xta,Klebanov:2022syt}. There   it was  proposed that the $M(3,10)$ and $M(3,8)$ minimal models are described by different fixed points of the GL action
\be
S =\int d^dx \bigg( \frac{1}{2}( \partial_\mu\phi)^2+\frac{1}{2}(\partial_\mu \sigma  )^2+\frac{g_1}{2}\sigma\phi^2+\frac{g_2}{6}\sigma^3 \bigg) \, ,  \label{GLprop}
\ee
and that $M(3,10)$, when deformed by the relevant operator  $\phi_{1,7}$, flows to the $M(3,8)$. 
We refer to \cite{Klebanov:2022syt,Katsevich:2024jgq} for more details and consistency checks. Here we briefly summarise the logic for the identification of the two CFTs and the RG-flow connecting them:
\begin{enumerate}
\item[1.-] First of all, resummations of perturbative $\epsilon$-expansion results are consistent with the identification of the GL action \reef{GLprop} with the $M(3,8)$.  
\item[2.-] After performing the field redefinition $\phi_1=(\sigma+\phi)/\sqrt{2}$, $\phi_2=(\sigma-\phi)/\sqrt{2}$, and    setting $g_1=g_2$,
it is revealed that  the  action \reef{GLprop} describes a pair of Lee-Yang effective actions \reef{GLLY} with equal couplings. At the same time it is known that the ($D_6$ modular invariant) $M(3,10)$ is given by the direct product of two $M(2,5)$, i.e. $M(3,10)_{D_6}=M(2,5)\otimes M(2,5)$~\cite{Kausch:1996vq,Quella:2006de} -- see appendix~\ref{m310} for further details. This completes the identification of the two CFTs at the endpoints of the flow.
\item[3.-]  As for the RG-flow, the field  $\phi_{1,7}\in M(3,10)_{D_6}$ is identified with the GL field $i(\phi_1^2\phi_2+\phi_2^2\phi_1)$, suggesting that $M(3,10)+\phi_{1,7}\rightarrow M(3,8)$.~\footnote{In the GL description $\mathcal{PT}$ acts as $i\to -i, \phi\to \phi$ and $\sigma \to -\sigma$, thus a factor of $i$ makes  $\phi_{1,7}$  even under  $\mathcal{PT}$.}  Indeed this perturbation \emph{undoes} the $g_1=g_2$  done in the GL indentification of $M(3,10)$ and allows for the action to reach the critical point corresponding to the $M(3,8)$.
This GL  RG-flow is then further supported by a number of consistency checks in $d=2$, such as: its consistency with  the  (generalised) $c$-theorem~\cite{Castro-Alvaredo:2017udm},   global symmetries and ${\cal PT}$-symmetries, and anomaly matching for non-invertible topological lines~\cite{Nakayama:2024msv}.
\end{enumerate}
All in all, there is mounting   evidence for 
\be
M(3,10)\, + \, \phi_{1,7}\  \rightarrow \  M(3,8) \, .  \label{conj}
\ee

The purpose of this work is to provide more evidence for  \reef{conj} using Hamiltonian Truncation.
The possibility of performing this crosscheck  was originally raised in \cite{Klebanov:2022syt}.

The general idea of Hamiltonian Truncation (HT) is  simple, and  consists in diagonalising the finite dimensional Hamiltonian matrix
\be
H_{ij}\equiv  \delta_{ij}\,E_i   + V_{ij} \quad , \quad \quad E_i \leq E_T \label{trunc1}
\ee
where we have defined  $V_{ij}\equiv   \langle i |  V | j \rangle $, and   $\{|i\rangle\}$ are orthonormal eigenstates of the free or solvable Hamiltonian $H_0|i\rangle=E_i |i\rangle$.
There is a  numerical error in the HT approach due to the finite truncation energy cutoff $E_T$.
The  true   spectrum is  recovered only after performing the extrapolation $E_T\rightarrow \infty$.

We are interested in  determining the spectrum of QFT Hamiltonians  that are
defined as relevant deformations away from an ultraviolet CFT.  We  place the UV CFT on the finite volume circle of radius $R$, which ensures that the spectrum of the Hamiltonian is discrete. 
The potential is taken to be the integral of a local operator $V=g\int dx\,O(x)$. 
For  relevant perturbations $g$ has positive  mass dimension. 
The spectrum of the truncated Hamiltonian converges to the QFT spectrum  with a power like rate $g^2/E_T^{a}$ with a positive value of $a$.~\footnote{The value of $a$ may be estimated via a perturbative calculation~\cite{Rychkov:2014eea,Hogervorst:2014rta,Elias-Miro:2015bqk,Cohen:2021erm}.}  
 The more relevant  the perturbation $V$ is, the faster the convergence.
 This method is therefore complementary to conformal perturbation theory expansions of nearly marginal operators: it works best when the dimension of the perturbation is smallest. Clearly, the method also works best at weak coupling. However,  one can find the spectrum non-perturbatively at strong coupling by diagonalising \reef{trunc1}.

Yurov and Zamolodchikov \cite{Yurov:1991my} introduced the HT approach to study the RG-flow defined by deforming the  two-dimensional Lee-Yang CFT by its single relevant primary, and soon after the same approach was used to study deformations of the two-dimensional Tricritical Ising~\cite{Lassig:1990xy}. In this context `Hamiltonian Truncation' was termed Truncated Conformal Space Approach (TCSA). 
The TCSA approach has been generalised to higher dimensions \cite{Hogervorst:2014rta} and to  study  relevant deformations of UV divergent QFTs \cite{EliasMiro:2022pua,Delouche:2023wsl}~\footnote{
Other works have dealt with UV divergencies appearing at leading order, such as the $\log$ divergent vacuum of the $M(4,5)+\eps^\prime$ flow~\cite{Giokas:2011ix} and Schwinger Model~\cite{Ingoldby:2024fcy}, or the $\log$-divergent mass in the $\text{CFT}_\text{free}+\phi^2+\phi^4$ in the light-cone conformal truncation~\cite{Anand:2020qnp}. Refs.~\cite{Rutter:2018aog,Elias-Miro:2020qwz} explored the needed counter-terms for the three-dimensional  $\phi^4$  scalar field theory. }. It is this second generalisation that we will need in this work.

From the point of view of Hamiltonian Truncation, computing the RG-flow \reef{conj} is an interesting challenge. The operator $\phi_{1,7}$ leads to coupling constant renormalization up to third order in the coupling. To our knowledge this is a situation in which HT techniques have never been employed. The complexity resides in analytically  determining all the necessary counter-terms, and in implementing efficient numerics to reach the  strong coupling
regime. Therefore, computing this RG-flow marks a new milestone in the capabilities of HT methods.

The paper is organised as follows: in section~\ref{reviewHT} we review the HT method and the generalization to study RG-flows that require UV renormalization. In section~\ref{TH} we explain the particulars of this program for the $M(3,10)+\phi_{1,7}$ QFT. 
Finally, in section~\ref{NUM} we proceed to diagonalise the renormalized truncated Hamiltonian of this theory. 
We study the ${\cal PT}$-symmetry breaking phases of this RG-flow and provide  evidence that the IR is indeed described by the minimal model  $M(3,8)$.

\section{Review of Hamiltonian Truncation crafted for UV-divergent QFTs}
\label{reviewHT}

\subsection{Hamiltonian Truncation for Minimal Models}

In this section we adapt the general formulation of Hamiltonian Truncation (HT) presented in the introduction to the case of QFTs defined by deforming $d=2$ minimal models on the cylinder $\mathbb{R} \times S_R^1$. This section is mostly a review, and serves to define our notation. 

We use cylinder coordinates that are  defined as $(\tau, x) = R(\log r, \theta)$, where  $(r,\theta)$  are the polar coordinates on the  $\mathbb{R}^2$-plane. Here
  $x \in [0,2\pi R)$ is the compactified spatial coordinate, while  $\tau \in \mathbb{R}$ is the time along the non-compact  cylinder direction. 
  The CFT Hamiltonian $H_{\text{CFT}}$ on this geometry can be written in terms of the Virasoro generators of the CFT on the plane:
\beq \label{eq:CFT-Hamiltonian} 
H_{\text{CFT}} = \frac{1}{R}\left( L_0 + \bar{L}_0 -\frac{c}{12}\right),
\eeq
where $c$ is the central charge. The full QFT Hamiltonian is defined by adding to the UV Hamiltonian relevant operators. In our particular case, we consider
\beq \label{eq:QFT-Hamiltonian}
H_\text{Bare}= H^{M(3,10)}_{\text{CFT}} +  \frac{1}{2\pi} \int_0^{2\pi R} dx  \, \big( \, g_3\phi_{3}(0,x) +i g_5\phi_{5}(0,x)+ g_7\phi_{7}(0,x)   \big)   \, ,
\eeq
where the $g_i$'s are dimensionful coupling constants of dimension $2-\Delta_i$ and the $\phi_{i}$'s are the $M(3,10)$ $\bZ_2$-even relevant primary scalar fields of dimension $\Delta_i$, evaluated at zero cylinder time. In table~\ref{tab:M310D6scal} we provide a summary of the scalar primaries in this theory.
Here, when there is no ambiguity,  we use the notation  $\phi_i\equiv \phi_{1,i}$,
  denote the OPE coefficient of three such operators  $C_{\phi_{1,i},\phi_{1,j}}^{\phi_{1,k}} \equiv C_{ij}^k$,
  and their  dimension simply by $\Delta_i$, with $i=\{3,5,7,9\}$.

The operator $H_\text{Bare}$ is $\mathcal{PT}$-symmetric. At weak coupling $g_i R^{2-\Delta_i} \ll 1$, its spectrum is real. However at strong coupling, the $\mathcal{PT}$ symmetry may be in a \emph{broken} phase, defined by the presence of a number of eigenvalues appearing in   complex conjugate pairs. We discuss the phases of the theory in Sec.~\ref{NUM}.
A number of recent  interesting works  have studied  $\mathcal{PT}$ broken/unbroken phases using HT:  Refs.~\cite{Lencses:2022ira,Lencses:2023evr} studied multicritical versions of the  Lee-Yang edge singularity,  and  Refs.~\cite{Xu:2022mmw,Xu:2023nke} precisely determined properties of the Lee-Yang edge singularity associated to the Ising  in a purely imaginary  magnetic field, such as the Lee-Yang EFT, and S-matrix. See also  our preceding study of the $M(3,7)+\phi_{1,5}$ flow  \cite{Delouche:2023wsl}.

The basis of states that we use to compute the matrix elements of the Hamiltonian are made up of a  string of Virasoro modes acting on primary states:
\beq \label{eq:basis-state-Virasoro}
\ket{\psi} = L_{-n_1}\ldots L_{-n_k}\bar{L}_{-m_1}\ldots \bar{L}_{-m_l}\ket{h}, \quad n_1\geq \ldots \geq n_k>0,m_1\geq \ldots \geq m_l >0,
\eeq
with $\ket{h} = \phi_{r,s}(0,0)\ket{0}$, where $(r,s)$ indices refer to the Kac table position.~\footnote{
In this work we consider  a subsector, closed under OPE, which contains scalar primaries only. Thus here the fields $\phi_{r,s}$ have the same holomorphic $h$ and anti-holomorphic $\bar h =h$ dimension, and total scaling dimension $2h$. 
More generally,  non-diagonal  modular invariant theories contain sectors with spinning primaries in their spectrum.
} 
The Hamiltonian in \eqref{eq:QFT-Hamiltonian} conserves spatial momentum,  and 
we choose to work in the zero momentum sector, where $\sum_{i=1}^k n_i = \sum_{j=1}^l m_j$.
 These states  are eigenstates  of the CFT Hamiltonian \eqref{eq:CFT-Hamiltonian}, with eigenvalue given by  $\frac{1}{R}(2h + 2\sum_{i}n_i -c/12) = \frac{1}{R}(\Delta_\psi -c/12)$.
Then, the matrix elements of a deformation term $V_j \equiv \frac{g_j}{2\pi} \int_0^{2\pi R} dx \phi_{\Delta_j}(0,x)$ between two states of the form \eqref{eq:basis-state-Virasoro} are given by
\beq \label{eq:V-matrix-elements}
\langle \psi_f| V_j | \psi_i \rangle =  g_j R \langle \psi_f| \phi_{\Delta_j}(0,0) | \psi_{i}\rangle =  g_j R^{1-\Delta_j} \tilde{C}_{fi}^j.
\eeq 
where  $\tilde{C}_{fi}^j$ refers to the dimensionless three point function  $ \langle \mathcal{O}_f(\infty) \phi_{\Delta_j}(1) \mathcal{O}_i(0)\rangle$, where $\phi_{\Delta_j}(1)$ is the plane primary field evaluated at an arbitrary point on the unit circle of the plane. 
The coefficients  $\tilde{C}_{fi}^j$ are completely fixed by conformal symmetry in terms of the CFT data $\{ \Delta_i, C^{i}_{jk}\}$ and can be computed numerically. When $\mathcal{O}_f, \mathcal{O}_i$ are primary operators, $\tilde{C}_{fi}^j$ is of course simply the OPE coefficient $C_{fi}^j$.
The three-point functions can be factorised into holomorphic and anti-holomorphic parts.~\footnote{We calculate the two parts recursively before combining them. We developed Mathematica and Python codes to compute and crosscheck our numerical results, which are available upon request.}

All in all, the matrix elements of the  Hamiltonian \eqref{eq:QFT-Hamiltonian} between our basis states are given by:
\beq \label{eq:Hamiltonian-matrix-elements}
\langle \psi_f|H|\psi_i\rangle = \delta_{\Delta_f,\Delta_i}\langle\psi_f|\psi_i\rangle \frac{1}{R}\left(\Delta_i-\frac{c}{12}\right) +   \sum_{j} g_j R^{1-\Delta_j} \tilde{C}_{fi}^j.
\eeq
Finally we remark that the basis $\{|\psi_i\rangle\}$ we'll be working with is not orthonormal, 
and thus we deal with the generalized eigenvalue problem   $H\vec{v} = E G \vec{v}$, 
where $G_{ij}=\langle \psi_i | \psi_j \rangle$ is the non-singular~\footnote{We construct our basis iteratively, adding Virasoro descendants with the structure indicated in  \reef{eq:basis-state-Virasoro} one by one and accepting the new state only if the Gram matrix has no zero eigenvalues. This ensures our truncated basis is free of Virasoro null states with zero norm.} Gram matrix. 

\subsection{$M(3,10)$ Minimal Model}

For future reference we provide here  further details about  the $M(3,10)$ CFT. 
The scalar primaries and its transformation properties are given in table~\ref{tab:M310D6}.
\begin{table}[h]
    \centering
    \begin{tabular}{l l l l l l l l l l l }
 $M(3,10)_{D_6}$    &  $\phi_{1,1}$ \phantom{--} & $\phi_{1,3}$   \phantom{--}  & $\phi_{1,5}$   \phantom{--} & $\psi_{1,5}$    \phantom{--}  & $\phi_{1,7}$     \phantom{--} & $\phi_{1,9}$     \phantom{\Big|} \\
\hline
\hline
    $\Delta$ &  0  & -4/5 & -2/5 &-2/5 & 6/5 & 4   \phantom{\Big|}\\
\hline
$\bZ_2$ & + & + & + & --  & + & +  \phantom{\Big|} \\
\hline
$\mathcal{PT}$ & + & + &  -- & +& + & +  \phantom{\Big|} \\
\hline
    \end{tabular}
    \caption{The scalar  primaries of the $M(3,10)$ minimal model ($D_6$ modular invariant), along with their  dimensions, and transformation properties under a global $\bZ_2$-symmetry  and $\mathcal{PT}$-symmetry.}
    \label{tab:M310D6scal}
\end{table}

\noindent The  fusion rules of the $\bZ_2$-even primaries are given by
   \bea
 \phi_{1,3}\times \phi_{1,3} & \sim 1+\phi_{13}+i\phi_{1,5}   &    \phi_{1,3}\times  \phi_{1,5}  & \sim i \phi_{1,3}+\phi_{1,5}+i\phi_{1,7} 	\\
 \phi_{1,5}\times \phi_{1,5} & \sim 1+\phi_{13}+i\phi_{1,5} +\phi_{17}+\phi_{19} & \phi_{1,3}\times  \phi_{1,7}  & \sim i\phi_{1,5}+\phi_{1,7}+\phi_{1,9} \\
  \phi_{1,7}\times \phi_{1,7} & \sim 1+\phi_{13}+i\phi_{1,5}  & \phi_{1,3}\times \phi_{1,9}& \sim \phi_{1,7}   \\
  \phi_{1,9}\times \phi_{1,9} & \sim 1  & \phi_{1,5}\times \phi_{1,7}& \sim  i \phi_{1,3}+\phi_{1,5}+i\phi_{1,7}   \\
 \phi_{1,5}\times \phi_{1,9}& \sim \phi_{1,5}   & \phi_{1,7}\times \phi_{1,9}& \sim \phi_{1,3}   
 \label{m310opescal}
 \eea
 In appendix~\ref{m310} we provide the values of the OPE coefficients $C_{ij}^k $ and all the necessary  details about
 this CFT. We also perform a detailed matching between $M(3,10)_{D_6}$ and $M(2,5)\otimes M(2,5)$.

  \subsection{Renormalization in Perturbation Theory}
\label{ptrenorm}

In this section we implement renormalization of the theory \reef{eq:QFT-Hamiltonian} in perturbation theory.
We first determine the structure of UV divergences that arise by computing corrections to the spectrum in conformal perturbation theory. Once the UV divergences are identified and regulated, we introduce counterterms necessary to absorb the divergences.

Perturbative corrections to the spectrum take the form of integrals over $n$-point correlation functions in the $M(3,10)$ CFT. These correlation functions have power-law singularities when two or more operators are brought to the same point. If there are any singularities strong enough to cause the overall integral to diverge, they should be interpreted as UV divergences.  For more details, see for example Ref.~\cite{EliasMiro:2021aof}. 

The underlying QFT is defined on the cylinder $\mathbb{R}\times S_R^1$, but correlation functions are often simpler to analyse on the plane. To facilitate conformal perturbation theory, we employ a conformal Weyl transformation to map integrated correlation functions to the plane from the cylinder. On the plane $\mathbb{R}^2$, we use complex coordinates $z_i$ or $w_j$, with all operators or correlation functions expressed in terms of these coordinates understood as defined on the plane. 

UV divergences first appear at the second order of perturbation theory in the coupling $g_7$. Corrections to energy differences between excited primary states and the vacuum state $\delta {\cal E}_\psi\equiv (E_\psi-E_0) R$ at this order are given by
\be
\delta {\cal E}_\psi^{(2)}= -  \frac{g_7^2R^{4-2\Delta_7}}{2\pi}    \int_{\substack{|z|\leq 1\\|1-z|>\eps}}\frac{d^2z}{|z|^{2-\Delta_7}} \bigg( \langle \psi(\infty)  \phi_7(1)\phi_7(z) \psi(0) \rangle_c -(C_{\psi\psi}^\phi)^2|z|^{-\Delta_7}\bigg)  \, , \label{entwo}
\ee
for primary operators $\psi(z)$~\cite{EliasMiro:2021aof}, where we use the notation $\psi(\infty)=\lim_{s\rightarrow \infty} |s|^{2\Delta_n}\psi(s)$. The four-point function is connected with respect to the four fields, as indicated by the subscript $c$. The integral in \reef{entwo} has no divergences coming from regions where the two $\phi_7$ fields are far apart, indicating that it is IR finite. There is however a UV divergence when the $\phi_7$ fields approach one another for $z\rightarrow 1$, which must be regulated. We do this by cutting out a ball of radius $\epsilon$ around this point from the domain of integration.

The UV divergences can be characterized in more detail by taking the $\phi_{1,7}\times\phi_{1,7}$ OPE in the four-point function (we use holomorphic variables $(z,\bar z)$ as coordinates on the plane):
\bea
\phi_7(z,\bar z) \phi_7(0)=|z|^{-2\Delta_7}\big[\mathds{1}+\cdots \big] &+ |z|^{-2\Delta_7+\Delta_3} C_{77}^3\big[\phi_3(0) + \frac{z}{2}  (L_{-1}\phi_3)(0)  + \frac{\bar z}{2} (\bar L_{-1}\phi_3)(0) +\cdots \big] \nonumber\\ &+ |z|^{-2\Delta_7+\Delta_5} C_{77}^5\big[\phi_5(0) + \cdots\big] 
\label{eq77ope}
\eea
and retaining only terms that generate non-integrable singularities in the $\epsilon\rightarrow0$ limit. The resulting UV divergences have the form
\bea
\delta {\cal E}_\psi^{(2)}\sim   \eps^{2-2\Delta_{7}+\Delta_5}  (\phi_{5})_\psi +    \eps^{2-2\Delta_{7}+\Delta_3} (\phi_{3})_\psi
+ \eps^{2-2\Delta_{7}+\Delta_3+1} (\phi_{3})_\psi\,. \label{sedor}
\eea
The notation $({\cal O})_{\psi}$ is shorthand for the plane correlation function $\langle\psi(\infty){\cal O}(1)\psi(0)\rangle$. We see that the OPE constrains the UV divergences to be proportional to local operators.

In \reef{eq77ope}, we have included the contribution from the leading descendant of $\phi_{1,3}$. In writing \reef{sedor}, we have used the fact that diagonal matrix elements of $zL_{-1}\phi_{1,3}+\bar{z}\bar{L}_{-1}\phi_{1,3}$ are proportional to those of $\phi_{1,3}$. We interpret the UV divergence coming from the descendant field in \reef{eq77ope} as a subleading UV divergence proportional to $\phi_{1,3}$.

At third order we encounter new UV divergences. Schematically, corrections to energies depend on integrated correlators with three insertions of $\phi_7$. When the three operators are close together, we can use the following `generalized'  OPE~\cite{EliasMiro:2022pua} to find the short-distance singularities
\bea
\delta{\cal E}_\psi^{(3)} &\sim \int_\eps  \langle \psi(\infty) \phi_7(z_1) \phi_7(z_2) \phi_7(1) \psi(0) \rangle_c \, ,\nonumber  \\ &\sim \sum_{\cal O} \langle \psi(\infty) {\cal O}(1) \psi(0) \rangle \int_\eps  \langle {\cal O} (\infty)\phi_7(z_1)\phi_7(z_2)\phi_7(1) \rangle_c\,.  \label{tresEN}
\eea
We ignore contributions from regions where only two or none of the $\phi_7$ fields approach each other; they contain sub-divergences (the same divergences we already encountered at second order \reef{sedor}) and finite terms. 

The coordinates $z_1$ and $z_2$ are integrated over in \reef{tresEN}, and $\eps$-balls around the singular points are removed when any of the $\phi_7$ fields approach one other, i.e. we impose $|1-z_i|>\eps$ and $|z_1-z_2|>\eps |z_2|$, see Ref.~\cite{EliasMiro:2022pua}. By considering how the integral on the second line of \reef{tresEN} transforms under scale transformations, we find that it is only UV divergent when ${\cal O}$ is either $\phi_3$ or $\phi_5$. Thus we are lead to 
\be
\delta{\cal E}^{(3)}_\psi \sim \eps^{4-3\Delta_7+\Delta_5} (\phi_{5})_\psi+ \eps^{4-3\Delta_7+\Delta_3} (\phi_{3})_\psi\,.
\label{tredor}
\ee
Therefore at third order, counter-terms are needed to account for a log-divergence proportional to  $\phi_5$ and a power-like divergence proportional to $\phi_3$. 

Going to fourth order, we find that the relevant integrated correlation functions are UV finite in the region where all four $\phi_7$ fields come together at once, according to power-counting using a local short-distance regulator. Therefore there is no need to introduce counterterms at fourth order or higher in perturbation theory.  We also note that, by a simple power counting argument analogous to the one just presented, there are no UV divergences due to collisions of $\phi_7$ against  $\phi_{3}$ and $\phi_5$ or due to $\phi_{3}$ and $\phi_5$ among themselves.

\subsubsection*{Counterterms at Second Order}

To absorb UV divergences in \reef{sedor} that appear at second order perturbation theory, we add the following interactions to the Hamiltonian
\be
{\cal V}_{c.t.}^{(2)}(0,x)  = \left(\frac{g_7}{2\pi}\right)^{2} \sum_{{\cal O}=\mathds{1},\phi_3,\phi_5}\lambda^{(2)}_{\cal O}(\eps)  \, {\cal O}(0,x) +
\left(\frac{g_7}{2\pi}\right)^{2} \gamma^{(2)}(\eps)  \, \frac{2}{5R}\, \phi_3(0,x)  \, , \label{density-cts}
\ee
where we use curly letters to denote operator densities on the cylinder so that $\int_0^{2\pi R}dx{\cal V}_\text{c.t.}(0,x)=V_\text{c.t.}$. The sum in the first term of \reef{density-cts} runs over primary fields, and includes the identity operator. The counterterm proportional to the identity modifies the vacuum energy but has no effect on energy differences. The second term comes from the descendant, which has matrix elements proportional to $\phi_3$. The functions $\lambda^{(2)}_{\cal O}(\eps) $ are second order counterterms, and specify our choice of renormalization scheme. They are
\be
\lambda_{\cal O}^{(2)}(\eps) =  R^{2-2\Delta_7+\Delta_{\cal O}} \int_{\substack{|z| \leq 1 \\ |1-z|>\eps}} d^2z |z|^{\Delta_7-2} \langle {\cal O}(\infty) \phi_{7}(1)\phi_{7}(z)\rangle\,, \label{ct2}
\ee
where the operators and integrals  are on the $\mathbb{R}^2$-plane, and ${\cal O}=\{\mathds{1},\phi_3,\phi_5\}$. The counterterm for the divergence that comes from the descendant $\varphi(z,\bar{z}) \equiv z(L_{-1}\phi_3)(z,\bar{z})$ and its antiholomorphic counterpart $\bar{\varphi}(z,\bar{z}) \equiv \bar{z}(\bar{L}_{-1}\phi_3)(z,\bar{z})$ is
\begin{align}
\gamma^{(2)}(\eps) = R^{2-2\Delta_7+\Delta_3+1} \int_{\substack{|z| \leq 1 \\ |1-z|>\eps}} d^2z |z|^{\Delta_7-2} \left\{\langle \varphi(\infty) \phi_{7}(1)\phi_{7}(z)\rangle + \langle \bar{\varphi}(\infty) \phi_{7}(1)\phi_{7}(z)\rangle \right\}\label{ct2des}\,,
\end{align}
where the correlation function involving the descendant at infinity takes the form $\langle \varphi(\infty)\phi_7(1)\phi_7(z,\bar{z})\rangle= C^3_{77}(z-1)/(2|1-z|^{16/5})$, and similarly for the antiholomorphic descendant.

\subsubsection*{Counterterms at Third Order}

The third order counterterms are given by
\be
{\cal V}_{c.t.}^{(3)}  = \left(\frac{g_7}{2\pi}\right)^{3} \sum_{{\cal O}=\phi_3,\phi_5}\lambda^{(3)}_{\cal O}(\eps)  \, {\cal O}(0,x) \, . \label{eq:vct3}
\ee
where the third-order counterterm of the $\phi_3$ operator is given by 
\be
\lambda_{{\cal O}}^{(3)}(\epsilon)=  -R^{4+\Delta_{\cal O}-3\Delta_{7}}\int\limits_{\substack{R,\,|1-z_i|>\epsilon\\|z_1-z_2|>\epsilon|z_2|}} \prod_{i=1}^2d^2z_i |z_i|^{\Delta_{7}-2}\big[  \, \langle {\cal O }(\infty)\phi_{7}(1)\phi_{7}(z_2)\phi_{7}(z_1)\rangle -   F_{\text{sub-divs}}^{{\cal O}} \, \big] \, ,  \label{eq:ct-3order-phi3} 
\ee
where $R$ indicates radial ordering $0\le|z_1|\le|z_2|\le1$. The counterterm should only cancel the divergence that arises from all three  $\phi_7$ insertions approaching each other. In particular, any divergences that correspond to pairs of points coming close with the third far away must be removed, because they are already cancelled by contributions from second order counterterms. We therefore subtract from the four-point function in \reef{eq:ct-3order-phi3} all singularities of this type.

In the case of the ${\cal O}=\phi_{3}$ counterterm, we subtract the singularities at $z_1\rightarrow z_2$, $z_1\rightarrow 1$ and $z_2\rightarrow 1$ that come from the $\phi_7\times\phi_7\sim\phi_5$ term in the OPE. The terms we need to subtract are
\be
F_\text{sub-divs}^{\phi_3}= C^5_{77}C^5_{37}\bigg(\frac{1}{|1-z_1|^{8/5}|1-z_2|^{14/5}}+\frac{1}{|1-z_1|^{14/5}|1-z_2|^{8/5}}+\frac{1}{|z_2-z_1|^{14/5}|1-z_2|^{8/5}} \bigg)\,.
\ee  
In the case of the ${\cal O}=\phi_{5}$ counterterm, we subtract the singularities that come from the terms in the OPE where two $\phi_7s$ fuse to form $\phi_3$, $\phi_5$ and the leading descendants $L_{-1}\phi_3$ and $\bar{L}_{-1}\phi_3$. Together, these terms are
\bea
F_\text{sub-divs}^{\phi_5} &= C^5_{77}C^5_{57}\left(\frac{1}{|1-z_1|^{6/5}|1-z_2|^{14/5}}+\frac{1}{|1-z_1|^{14/5}|1-z_2|^{6/5}}+\frac{1}{|z_2-z_1|^{14/5}|1-z_2|^{6/5}}\right)  \nonumber \\
& + C^3_{77}C^3_{57}\left(\frac{1}{|1-z_1|^{4/5}|1-z_2|^{16/5}}+\frac{1}{|1-z_1|^{16/5}|1-z_2|^{4/5}}+\frac{1}{|z_2-z_1|^{16/5}|1-z_2|^{4/5}}\right) \nonumber  \\
& -\frac{C^3_{77}C^3_{57}}{5}\left(\frac{(z_2-1)(1-\bar{z}_1)}{|1-z_1|^{14/5}|1-z_2|^{16/5}}+\frac{(z_1-1)(1-\bar{z}_2)}{|1-z_1|^{16/5}|1-z_2|^{14/5}}+\frac{(z_1-z_2)(\bar{z}_2-1)}{|z_2-z_1|^{16/5}|1-z_2|^{14/5}}\right) \nonumber \\
&  -\frac{C^3_{77}C^3_{57}}{5}\left(\frac{(\bar{z}_2-1)(1-z_1)}{|1-z_1|^{14/5}|1-z_2|^{16/5}}+\frac{(\bar{z}_1-1)(1-z_2)}{|1-z_1|^{16/5}|1-z_2|^{14/5}}+\frac{(\bar{z}_1-\bar{z}_2)(z_2-1)}{|z_2-z_1|^{16/5}|1-z_2|^{14/5}}\right)
\eea
In calculating the terms from the descendants, we have used the three-point function \newline$\langle \phi_5(\infty) (L_{-1}\phi_3)(w,\bar{w})\phi_7(z,\bar{z}) \rangle=(-2/5) C_{37}^5 (\bar w-\bar z)/|w-z|^{14/5}$, and similarly for $\bar{L}_{-1}\phi_3$.

In order to fully determine the third order counterterms \reef{eq:ct-3order-phi3}, we need to compute a couple of four-point functions: $\langle \phi_3 \phi_7  \phi_7  \phi_7 \rangle$ and $\langle \phi_5  \phi_7  \phi_7  \phi_7 \rangle$. We discuss their calculation in the next section. There we show that, for the $K_\text{eff}$ operator we only need the expansion of  $\langle \phi_{i=3,5}(\infty)  \phi_7(1)  \phi_7(z) \phi_7 (0)\rangle$ around $z=0$. These series expansions can be obtained by solving the BPZ equations  (full details in appendix~\ref{BPZapp}). 

As a crosscheck we have computed series expansions for these four-point functions using two additional methods: Firstly we use the formulation in terms of the direct product $M(2,5)\otimes M(2,5)\simeq M(3,10)_{D_6}$ and the known four-point function of the $M(2,5)$ (appendix~\ref{4ptDP}), and secondly we employ the coulomb gas formalism (appendix~\ref{CGapp}).

\section{Applying Hamiltonian Truncation to the $M(3,10)+\phi_{1,7}$  RG-flow}
 \label{TH}
 
  \subsection{Renormalization in Hamiltonian Truncation }
 
 \label{rTH}
 
 Imposing a cutoff $\Delta_T$ on the Hilbert space acts as a UV regulator. However this regulator doesn't preserve Lorentz invariance or act locally, so extra non-local counterterms can be needed for renormalization~\cite{EliasMiro:2021aof}. To find all the counterterms necessary for Hamiltonian Truncation, we proceed via the method advocated in \cite{EliasMiro:2022pua}, which we briefly review. This method contains three steps:
 \begin{enumerate}
 	\item[1)]  First, perform renormalization using a standard, local regulator. For the $H_\text{CFT}^{M(3,10)}+\phi_{1,7}$ theory, this step is done in Section~\ref{ptrenorm}. There, we introduce a short distance regulator $\epsilon$ and a finite set of local counterterms to remove UV divergences from integrated correlation functions of the deformation operators. For any super-renormalizable theory, all the necessary counterterms can be found by considering finitely many orders in perturbation theory. Going up to third order in the coupling $g_7$ is sufficient to fully renormalize the theory in our case. Afterwards, the new Hamiltonian takes the form $H=H_\text{Bare}  +V_{\text{c.t}.}(\eps) $, where $V_{\text{c.t}.}(\eps)$ includes the counterterms, i.e. local operators multiplied by  functions of $\eps$ that diverge  as $\eps \rightarrow 0$.
 	
 	\item[2)] Next, we introduce an Effective Hamiltonian for the renormalized QFT, constructed to have the following properties: it must have finite dimensionality and act in a truncated low-energy Hilbert space, while also having the same low-energy spectrum of the full renormalized QFT up to a fixed order in perturbation theory. We take our truncated Hilbert space to be the space spanned by the $M(3,10)$ CFT Hamiltonian eigenstates with eigenvalues such that $ \Delta_i \leq \Delta_T$ and build our Effective Hamiltonian to match the exact spectrum to third order in $g_7$. Schematically, we are led to the following Effective Hamiltonian 
 	\be
 	\delta_{ij}\,\Delta_i + V_{ij}+ V_{\text{c.t}.}(\eps)_{ ij} + H_{\text{eff},2}(\eps)_{ij}  + H_{\text{eff},3} (\eps)_{ij}  \quad , \quad \quad \Delta_i,\Delta_j \leq \Delta_T\,.  \label{trunc2}
 	\ee 
 	The Effective Hamiltonian takes into account the effect of high energy states, $\Delta_i>\Delta_T$, on low-energy physics. In this respect it is analogous to a Wilsonian effective action, where the physical effects of modes above the cutoff are accounted for by adjusting the coefficients in an action for field modes below the cutoff.
 	
 	\item[3)] Finally, we take the $\eps\rightarrow 0$ limit of the matrix elements of the Effective Hamiltonian in \reef{trunc2}:
 	\be
 	K_\text{eff}   \equiv \lim_{\eps\rightarrow 0}\left[V_{\text{c.t.}}(\eps) + H_{\text{eff},2}(\eps) + H_{\text{eff},3} (\eps) \right]\,,
 	\label{kdef}
 	\ee
 	to obtain the renormalized Hamiltonian
 	\be
 	H_{ij} \equiv  \delta_{ij}\Delta_i\, + V_{ij} + (K_{\text{eff}})_{ij}\,. \quad \quad \Delta_i,\Delta_j \leq \Delta_T \label{hamct}
 	\ee 
 	In the theory at hand $K_\text{eff}=K_\text{eff,2}+K_\text{eff,3}$, where numbers in subscripts indicate the order in the coupling $g_7$.
 \end{enumerate}

 Crucially, the limit in  \reef{kdef} is finite. Let us provide an intuitive argument for why this claim is true at  second order in the coupling $g_7$: schematically $(V_{c.t.}(\epsilon))_{ij}\sim\sum_{E_k,\,\epsilon}V_{ik}E_k^{-1}V_{kj}$, where the sum is over all the states in the theory. The large-$k$ terms get modified more strongly by the short distance regulator $\epsilon$, ensuring the infinite sum is finite. Adding to the last expression $(H_\text{eff,2}(\eps))_{ij}\sim \sum_{E_k>E_T, \eps} V_{ik}\frac{1}{E_k} V_{kj}$. and removing the local regulator, $\eps\rightarrow0$ then leads to the $K_\text{eff}^{(2)}$ operator whose matrix elements are explicitly finite, since they are given in terms of a sum over low energy states. Note however that the $K_\text{eff}$ operator \emph{does} diverge in $\Delta_T$ precisely to account for the UV divergences of the $H_\text{Bare}$ theory. The operator $K_\text{eff}$ is non-local and cannot be written as the integral of a local density.
 
 Another feature of the method is that in \emph{step 1}, the scheme gets defined, and this scheme choice is unmodified in \emph{step 2}. For a more detailed review of the method, see Ref.~\cite{Delouche:2023wsl}.
 We have included table~\ref{tab:flows-summary} to summarise the models that we have studied following the renormalization procedure outlined here. The counterterms included in $K_{\text{eff}}$ are indicated in schematic form. We have omitted their non-local structure and highlighted only the local part, that dominates for  large $\Delta_T$.  The leading scaling in $\Delta_T$ is a consequence of the local scaling in \reef{sedor} and \reef{tredor}.

 \begin{table}[t]
 	\centering
 	\begin{tabular}{l  l  l   l}
 		Flow & Bare Hamiltonian & Counterterms & \\[.2cm]
 		\hline \hline
 		1. Ising + $\eps$  & $H^{M(3,4)}_{\text{CFT}} + m \epsilon^{(1)}$   $\phantom{\Big|}$ & $ m^2 \log{\Delta_T}\, \mathbb{1}$  &  Ref.~\cite{Delouche:2023wsl}  \\[.1cm]
 		\hline
 		2. Tricrit.~Ising+ $\eps^\prime$  & $H^{M(4,5)}_{\text{CFT}} + g \epsilon^{\prime (6/5)}$   $\phantom{\Big|}$& $g^2 \Delta_T^{2/5} \, \mathbb{1}$  &   Ref.~\cite{Delouche:2023wsl} \\[.1cm]
 		\hline
 		3. $M(3,7)+\phi_{1,5}$  & $H^{M(3,7)}_{\text{CFT}} + g_5 \phi_{1,5}^{(8/7)}$   $\phantom{\Big|}$ & $g_5^2 \Delta_T^{2/7}\, \mathbb{1}+ g_5^2 \Delta_T^{4/7}\, \phi_{1,3}$ &   Ref.~\cite{Delouche:2023wsl} \\[.1cm]
 		\hline
 		\multirow{2}{*}{4. $M(3,10)+\phi_{1,7}$} &    \multirow{2}{*}{$H^{M(3,10)}_{\text{CFT}} +  g_7 \phi_{1,7}^{(6/5)}$   $\phantom{\Big|}$} & 
 		\footnotesize{$ g_7^2\, \big( \Delta_T^{2/5}\mathbb{1} + \Delta_T^{4/5}\phi_{1,5}+( \Delta_T^{6/5}+ \Delta_T^{1/5})\phi_{1,3}\big){+}$ }&  \multirow{2}{*}{\S~\ref{TH}-\ref{NUM}}\\
 		&     & 
 		\footnotesize{$ g_7^3\, \big( \log\Delta_T  \, \phi_{1,5}+ \Delta_T^{2/5}\phi_{1,3}\big)$} & \\
 		\hline
 	\end{tabular}
 	\caption{A summary of   strongly coupled RG-flows that we have  studied via Hamiltonian Truncation with the renormalisation method reviewed in Sec.~\ref{rTH}.  The dimension of the perturbing operator is indicated with a superscript.  }
 	\label{tab:flows-summary}
 \end{table}

\subsection{Calculation of the Effective Hamiltonian}

In the following, we apply the renormalization method from Section \ref{rTH} to the $M(3,10) + \phi_{1,7}$ QFT. This involves specifying the Effective Hamiltonian and computing the operator $K_\text{eff}$. This choice is not unique, and in this work we adopt the Schrieffer-Wolff Effective Hamiltonian \cite{SWtrans}, a widely used approach in the literature.

\subsubsection*{Second order}
Our calculation of the operator $K_\text{eff}$ at second order in perturbation theory closely follows the steps presented in appendix~A of Ref.~\cite{Delouche:2023wsl}. The Schrieffer-Wolff Hamiltonian at this order is given by  
\begin{align}
	\left(H_{\text{eff,}2}\right)_{fi} = \frac{1}{2}\sum_{E_h> E_T}\left(\frac{V_{fh}V_{hi}}{E_{fh}}+\frac{V_{fh}V_{hi}}{E_{ih}}\right)\,,
	\label{eq:sw2}
\end{align}
where $E_{ij}\equiv E_i-E_j$. Taking the interaction $V$ to be the $\phi_7$ field integrated over space, we can rewrite the sums in \reef{eq:sw2} in terms of matrix elements of the field
\be \label{eq:exact-heff2}
\frac{R}{2} \sum_{\Delta_h>\Delta_T}\frac{V_{fh}V_{hi}}{\Delta_{i}-\Delta_h} = -\frac{g_7^2}{4\pi} R^{3-2\Delta_7}  \oint_{|z|=1}\frac{dz}{2\pi i z}\int\limits_{\substack{0\leq |w| \leq 1\\|z-w|>\epsilon}}d^2w |w|^{\Delta_7 -2}\bra{f}\phi_7(z)\myline^i \phi_7(w)\ket{i}\, ,
\ee
where $\myline^i$ denotes a partial insertion of the identity in which only the contribution from high energy states $\Delta_h>\Delta^{T}_{7,i}\equiv \Delta_T - \Delta_7 - \Delta_i$ is retained. That is $\myline^i=\sum_{\Delta_h>\Delta^T_{7,i}}|h\rangle\langle h|$. Note also that this expression has UV divergences coming from the $w\rightarrow z$ region, which we regulate with the short-distance cutoff $\epsilon$. Taking the most singular terms in the OPE between two $\phi_7$ fields in \reef{eq:exact-heff2} yields
\bea
H_{\text{eff,}2}(\epsilon)_{fi}=&-\frac{g_7^2}{4\pi}R^{3-2\Delta_7}\sum\limits_{\mathcal{O}=\mathbb{1},\phi_3,
	\phi_5} (\calO)_{fi} \int\limits_{\substack{0\leq |w| \leq1\\ |1-w|> \epsilon}} \frac{d^2w}{ |w|^{2-\Delta_7}}\langle \mathcal{O}(\infty)\phi_7(1) \myline^i \phi_7(w) \rangle \nonumber  \\
& -\frac{g_7^2}{10\pi}R^{3-2\Delta_7}(\phi_3)_{fi} \int\limits_{\substack{0\leq |w| \leq1\\ |1-w|> \epsilon}} \frac{d^2w}{ |w|^{2-\Delta_7}}\langle \varphi(\infty)\phi_7(1) \myline^i \phi_7(w) \rangle + f\leftrightarrow i    \, ,
\eea
where we denote the cylinder matrix elements $\int_0^{2\pi R}dx\matrixel{f}{\calO(0,x)}{i}=2\pi R^{1-\Delta_\calO}\,(\calO)_{fi}$, and use the fact that matrix elements of the descendants $\varphi+\bar{\varphi}$ are proportional to $\phi_3$.

We add to the Effective Hamiltonian the counter-terms to second order from \reef{ct2} and \reef{ct2des}. Then we can safely take the limit $\eps\rightarrow 0$, leading to 
\be
(K_{\text{eff,2}})_{fi}= \frac{g_7^2 R^{3-2\Delta_7}}{2}\bigg[  \,  \alpha^{\mathds{1}}_{\Delta^T_{7,i}} \ (\mathds{1})_{fi}   \,  + \big[ \, \alpha^{3}_{\Delta^T_{7,i}}  +\beta^{3}_{\Delta^T_{7,i}}\,  \big] \ (\phi_3)_{fi}  \, +\,  \alpha^{5}_{\Delta^T_{7,i}} \ (\phi_5)_{fi} \, \bigg]+f \leftrightarrow i\,.  \label{keff2}
\ee
The expression for $K_\text{eff,2}$ is nonlocal, since the coefficients appearing in front of local operators also depend on the external state indices $f$ and $i$. The coefficients take the form
\be
\alpha^{s}_{\Delta^T_{7,i}}= C_{77}^{s}\sum_{n=0}^{2n\leq \Delta^T_{7,i}}\frac{u_n^{\Delta_7-\Delta_s/2}}{2n+6/5}
\quad \text{and }\quad 
\beta^{3}_{\Delta^T_{7,i}}=-\frac{2}{5} C_{77}^{3}\sum_{n=0}^{2n\leq \Delta^T_{7,i}}\frac{r_n^{\Delta_7-\Delta_3/2}r_n^{\Delta_7-\Delta_3/2-1}}{2n+6/5} \, , \label{coeff2}
\ee
where we have defined 
\be
r_n^\Delta\equiv \Gamma(n+\Delta)/(\Gamma(\Delta)n!) \quad \text{and}\quad u_n^\Delta\equiv ( r_n^\Delta)^2\,.
\ee
Further details on the calculation of $\beta^{3}_{\Delta^T_{7,i}}$ can be found in appendix~\ref{desct}. 

The sums in \reef{coeff2} run over finitely many states with energies below the cutoff $\D$, but they do present the correct UV divergences in the limit $\Delta_T \to \infty$. Indeed we have
\be
\alpha^{s}_{\Delta^T_{7,i}} \sim 
\Delta_T^{2\Delta_7-\Delta_s-2} \quad \text{and} \quad \beta^{3}_{\Delta^T_{7,i}} \sim  \Delta_T^{2\Delta_7-\Delta_3-3} \, , 
\ee
which follows from $r_n^\Delta\sim n^{\Delta-1}/\Gamma(\Delta)(1+O(1/n))$ as $n\to \infty$.

We have derived in \reef{keff2} the second order counterterms that are needed when a `$\Delta_T$' regulator is used. Instead of attempting to determine these nonlocal counterterms directly, we proceeded in an indirect way by first regulating the theory with a truly local cutoff and then constructing an Effective Hamiltonian, which delivers the desired counterterms when the local cutoff $\eps$ is taken arbitrarily small. 

\subsubsection*{Third order}

At third order in $g_7$, there are two distinct types of contributions to $H_{\text{eff},3}$. One that comes directly from the Schrieffer-Wolff Hamiltonian expanded to third order, and another which arises as an effective interaction between counterterms we added at second order and $V$, so that overall we have
\be
(H_{\text{eff},3})_{fi}= (\Delta H_3)_{fi}+(\Delta H_2)_{fi} \, .  \label{heff3eps}
\ee
The direct third order contribution is 
\be
(\Delta H_3)_{fi} = \frac{1}{2}\sum_{\substack{E_h> E_T\\E_h^\prime> E_T}}\frac{V_{fh}V_{hh^\prime}V_{h^\prime i}}{E_{ih}E_{ih^\prime}}-\frac{1}{2}\sum_{\substack{E_h> E_T\\E_l\le E_T}}\frac{V_{fl}V_{lh}V_{hi}}{E_{hi}E_{hl}} + f\leftrightarrow i\,,\label{heff3direct}
\ee
and the second order counterterm part is
\be
(\Delta H_2)_{fi} = \frac{1}{2}\sum_{E_h> E_T}\left(\frac{V_{fh}(V_\text{c.t,2})_{hi}}{E_{ih}}+\frac{(V_\text{c.t,2})_{fh}V_{hi}}{E_{ih}}\right) + f\leftrightarrow i\,.\label{heff3ct2}
\ee

Taking $V$ to be the spatial integral of $\phi_7$, the expression in \reef{heff3direct} can be rewritten in terms of matrix elements of this local operator. The most UV sensitive parts of this expression occur when all three $\phi_7$ matrix elements act at the same point. After using the generalised OPE to extract the singular behaviour from this region, we find
\bea
{\Delta H_3(\epsilon)}_{fi}=& \frac{g_7^3R^{5-3\Delta_7}}{8 \pi^2}\sum_{\mathcal{O}=\phi_3,\phi_5}\left(\calO\right)_{fi}\int_{\text{R},\epsilon}d^2z_1 d^2z_2|z_1|^{\Delta_7-2}|z_2|^{\Delta_7-2} \nonumber \\
& \left[\langle\mathcal{O}(\infty)\phi_7(1)\myline^i\phi_7(z_2)\myline^i\phi_7(z_1)\rangle-\langle\mathcal{O}(\infty)\phi_7(1)\myline^i\phi_7(z_2)\myline_i\phi_7(z_1)\rangle\right]+f \leftrightarrow i\,, \label{eq:heff3-3rd}
\eea
Where $R$ in the integral limit denotes radial ordering, and the $\epsilon$ subscript written there is shorthand for $|1-z_1|>\epsilon,|1-z_2|>\epsilon, |z_1-z_2|>\epsilon|z_2|$. This third order contribution includes a sum restricted to be over low energy states only, indicated by $\myline_i = \sum_{\Delta_l \leq \Delta^T_{7,i}}\ket{l}\bra{l}$.

To evaluate the integral expression in \reef{eq:heff3-3rd}, we make use of the following series expansions for the four point functions
\begin{align} \label{eq:def-BKQ}
	\langle \phi_{1,3}(\infty) \phi_{1,7}(1) \phi_{1,7}(z_2) \phi_{1,7}(z_1)\rangle &= C^3_{57}C^5_{77}\sum_{K,Q}^\infty B_{K,Q} \left(\frac{z_1}{z_2}\right)^{K}z_2^{Q-8/5} \times \text{c.c.}\,,\\
	\langle \phi_{1,5}(\infty)\phi_{1,7}(1)\phi_{1,7}(z_2)\phi_{1,7}(z_1)\rangle &=\sum_{s=3,5}C^s_{57}C^s_{77}\sum_{K,Q=0}^\infty B^s_{K,Q}\left(\frac{z_1}{z_2}\right)^K z_2^{Q-6/5+h_s}\times \text{c.c.} \, .
\end{align}
These expansions are useful, because upon integration over the $z_i$ variables, we can identify individual terms in the expansion with particular exchanged states~\cite{Delouche:2023wsl}. As a consequence, accounting for the incomplete sums over states (represented by e.g. $\myline^i$) in \reef{eq:heff3-3rd} is straightforward. The computation of the $B_{K,Q}$'s is detailed in appendix \ref{BPZapp}, where the series expansion of the various four-point functions are derived from solving BPZ equations. Their expressions are respectively given in Eq. \eqref{eq:5777-BKQ} and Eq. \eqref{eq:3777-BKQ}.

The contribution from second order counterterms can be rewritten to yield
\bea
{\Delta H_2(\epsilon)}_{fi}&=-\frac{1}{2}\sum\limits_{\substack{\mathcal{O}=\phi_3,\phi_5, \varphi}}\lambda^{(2)}_\mathcal{O}(\epsilon)R^{3-\Delta_\mathcal{O}-\Delta_7}\left(\frac{g_7}{2\pi}\right)^3\oint_{|z|=1}\frac{dz}{i z}\int\limits_{\substack{|w|\leq 1\\ |z-w|>\epsilon}}d^2w \label{eq:heff3-2nd} \\
&\left[|w|^{\Delta_7-2}\bra{f}\mathcal{O}(z)\myline \phi_7(w)\ket{i}+|w|^{\Delta_\mathcal{O}-2}\bra{f}\phi_7(z)\myline \mathcal{O}(w)\ket{i}\right]+f \leftrightarrow i\,. \nonumber
\eea
 When $\mathcal{O}=\varphi$ in the above, one should replace $\lambda_\mathcal{O}^{(2)}(\epsilon) \to \gamma^{(2)}(\epsilon)$. The contributions from both the holomorphic and antiholomorphic descendants are accounted for in the $\varphi$ term in Eq.~(\ref{eq:heff3-2nd}). We then simplify further by taking the OPE between the $\mathcal{O}$ and $\phi_7$ fields, and keeping only the terms proportional to the primaries $\phi_3$ and $\phi_5$. This is consistent with the truncation of the OPE that we are using in \eqref{eq:heff3-3rd}, and ensures that we keep only terms in OPEs that participate in the cancellation of UV divergences. Nevertheless, to improve convergence in $\Delta_T$, one could retain additional UV finite contributions both from \eqref{eq:heff3-3rd} and \eqref{eq:heff3-2nd}.

Finally, we add the third order counterterms shown in \reef{eq:vct3}. We then take the $\epsilon\rightarrow0$ limit, which is now finite and perform the spacetime integrations after expanding the correlators in series. After a lengthy calculation we are  lead to  
\bea \label{eq:Keff3-final}
(K_{\text{eff},3})_{fi}&= \kappa \, (\phi_{3})_{fi} \, C^5_{77}C^5_{37}\,  \Bigg\{      \sum_{K=0}\limits^{ 2K \leq  \Delta^T_{7,i}} \,\sum_{Q=0}\limits^{2Q \leq  \Delta^T_{5,i}}  \frac{(B_{K,Q})^2}{(2K+6/5)(2Q-2K-8/5)} + S_1 \Bigg\} \nonumber
\\
&+ \kappa \, (\phi_{5})_{fi} \, C^3_{77}C^3_{57}\,  \Bigg\{     \sum_{K=0}\limits^{ 2K \leq  \Delta^T_{7,i}} 	\, \sum_{Q=0}\limits^{2Q \leq  \Delta^T_{3,i}} \frac{(B_{K,Q}^{1,3})^2}{(2Q-2K-10/5)(2K+6/5)} + S_2 \Bigg\} 
\\
&+ \kappa \, (\phi_{5})_{fi} \, C^5_{77}C^5_{57}\,  \Bigg\{    \sum_{K=0}\limits^{ 2K \leq  \Delta^T_{7,i}} \, \sum_{Q=0}\limits^{2Q \leq  \Delta^T_{5,i}} \frac{(B_{K,Q}^{1,5})^2}{(2Q-2K-8/5)(2K+6/5)} +S_3 \Bigg\} +f \leftrightarrow i\,,\nonumber
\eea
where $\kappa\equiv   g_7^3R^{5-3\Delta_{7}}/2$, and $\Delta^T_{i,j}\equiv\D-\Delta_i-\Delta_j$. The sums above each include a finite number of terms, which grows with $\D$. These sums diverge as $\Delta_T\rightarrow\infty$ and capture the leading singularities of the correlators with 3 $\phi_7$ fields. The subtraction terms capture subdivergences (UV divergences that come from 2 out of 3 $\phi_7$ fields approaching each other) and are given by
\bea
S_1=& \sum\limits_{\substack{2K \geq 0\\Q=0}}^{2Q \leq  \Delta^T_{5,i}}\Big[\frac{u_K^{7/5}u_Q^{4/5}-(B_{K,Q})^2}{(2K+6/5)(2Q-2/5)}
\nonumber \\
+&\sum\limits_{\substack{2Q \geq 0\\K=0}}^{2K \leq \Delta_{7,i}^T }\Big[\frac{u_K^{4/5}u_Q^{7/5}}{(2K+6/5)(2Q+6/5)}-\frac{(B_{K,Q})^2}{(2Q-2K-8/5)(2K+6/5)}\Big]\,,\label{eq:s1}
\eea
\bea
S_2=&-\sum_{K=0}^\infty \frac{(B_{K,0}^{1,3})^2-u_K^{8/5}u_0^{2/5}}{(2K+6/5)(-4/5)}
+\sum\limits_{\substack{2K \geq 0\\Q=1}}^{2Q \leq  \Delta_{3,i}^T}\left[-\frac{(B_{K,Q}^{1,3})^2-u_K^{8/5}u_Q^{2/5}}{(2K+6/5)(2Q-4/5)}-\frac{2}{5}\frac{r_K^{8/5}r_K^{3/5}r_{Q-1}^{7/5}r_Q^{2/5}}{(2K+6/5)(2Q-4/5)}\right] \nonumber\\
&-\sum\limits_{\substack{2Q \geq 0\\K=0}}^{2K \leq \Delta_{7,i}^T}\left[\frac{(B_{K,Q}^{1,3})^2}{(2Q-2K-10/5)(2K+6/5)}{-}\frac{u_K^{2/5}u_Q^{8/5}}{(2K+6/5)(2Q+6/5)}{-}\frac{2}{5}\frac{r_K^{7/5}r_K^{2/5}r_Q^{8/5}r_Q^{3/5}}{(2K+6/5)(2Q+6/5)}\right]\,, \label{eq:s2}\\
S_3=&\sum\limits_{\substack{2K \geq 0\\Q=0}}^{2Q \leq  \Delta_{5,i}^T}\left[-\frac{(B_{K,Q}^{1,5})^2-u_K^{7/5}u_Q^{3/5}}{(2K+6/5)(2Q-2/5)}\right]\nonumber\\
+&\sum\limits_{\substack{2Q \geq 0\\K=0}}^{2K \leq \Delta_{7,i}^T  }\left[-\frac{(B_{K,Q}^{1,5})^2}{(2Q-2K-8/5)(2K+6/5)}+\frac{u_K^{3/5}u_Q^{7/5}}{(2K+6/5)(2Q+6/5)}\right]\,.\label{eq:s3}
\eea
The sums  in \reef{eq:s1} include an infinite number of terms, but nonetheless evaluate to finite values. The finiteness relies on delicate cancellations between components of the summand. 
The large $Q$, fixed $K$ sums in $S_2$ and $S_3$ individually diverge, but when combined as in \reef{eq:Keff3-final} lead to a finite result.
Next we discuss in more detail   the cancellation of divergences in the sums $S_1$, $S_2$ and  $S_3$.

On the first line of \reef{eq:s1}, the first term grows as $u_K^{7/5}\sim K^{4/5}$ for large $K$. This would cause the overall sum to diverge if there were not cancellations with $(B_{K,Q})^2$ terms, which have the same asymptotic growth for large $K$ (and fixed $Q$). These intricate cancellations can ultimately be traced back to cancellations between singularities in the correlators of \reef{eq:heff3-3rd} from regions where two fields come close, and second order counterterms that enter through \reef{eq:heff3-2nd}.

In \reef{eq:s2}, the summand contains contributions both from the second order $\phi_3$ counterterm, and from descendant $\varphi$ counterterms. Focusing on the top line for simplicity, we see that for large $K$, they grow as $u_K^{8/5}\sim K^{6/5}$ and $r_K^{8/5}r_K^{3/5}\sim K^{1/5}$ respectively, and cancel with leading and subleading growth of $(B^{1,3}_{K,Q})^2$ for large $K$. 
To ensure delicate cancellation of terms that are subleading at large $K$, we evaluate the  sums in \reef{eq:s2} using the prescription
\be
\sum_{2K\ge0}^{\infty} \frac{F(K)}{2K+X} \implies  \lim\limits_{y\rightarrow1^{-}} \lim\limits_{K_\text{max}\rightarrow \infty}  \sum_{2K\ge0}^{K_\text{max}}  \frac{F(K)}{2K+X} y^{2K+X}\,,
\ee
in this particular order. 
This is  motivated from the spacetime integral expressions for terms in $K_\text{eff,3}$. In which case, $y$ can be interpreted as a cutoff in a radial coordinate. The infinite sums over $Q$ are defined using the same prescription, which is important to ensure cancellations between the two blocks $S_2$ and $S_3$ at large $Q$.

We do not calculate $H_\text{eff,4}$ or $K_\text{eff,4}$ explicitly in this work. Using dimensional analysis, we can infer that the leading terms for large $\D$ will take the form 
\be
K_\text{eff,4}\sim g_7^4 \phi_{3} \Delta_T^{-2/5}+ g_7^4 \phi_{5} \Delta_T^{-4/5}\,. \label{keff4}
\ee
Furthermore there is also a term which goes as $ g_7^2 \phi_{1,5}\Delta_T^{-1/5}$ in $K_{\text{eff},2}$ from the exchange of the leading descendants of $\phi_{1,5}$, that we have not included. 
The powers that suppress all of these contributions have magnitude smaller than one, which suggests that the spectrum obtained from HT if these terms aren't included will converge to its infinite-$\D$ limit slowly. In the future it would be worthwhile to calculate these terms and incorporate them into the HT calculation.

\subsection{Final Result for the Renormalised  Hamiltonian}
 
 To summarise, the renormalized, truncated Hamiltonian for the $M(3,10)+\phi_{7}$ flow is given by
 \be
 H=H_\text{Bare} +   K_{\text{eff},2}+   K_{\text{eff},3} \, .  \label{todiag}
 \ee
 where the three terms are defined in \reef{eq:QFT-Hamiltonian}, \reef{keff2} and \reef{eq:Keff3-final}. In writing this Hamiltonian, a particular renormalization scheme has been selected. That scheme is defined by our choice of (local) counterterms. The counterterms we use are given in \reef{density-cts} and \reef{eq:vct3}.

\section{Numerical results}
 \label{NUM}

In this section we study the $M(3,10)$ perturbed by  any of its $\bZ_2$-even primaries. 
There are three renormalised couplings, $(g_3,g_5,g_7)$. We parametrise the space  via spherical coordinates
\bea
	g_3 & = \text{sign}(\sin\theta\cos\phi)\left|t\sin\theta\cos\phi\right|^{14/5}\,,\\
	g_5 & = \text{sign}(\sin\theta\sin\phi)\left|t\sin\theta\sin\phi\right|^{12/5}\,,\\
	g_7 & = \text{sign}(\cos\theta)\left|t\cos\theta\right|^{4/5}\,.
\eea
where exponents have been chosen so that $t$ has dimension $1$. As a result, physics only depends on the dimensionless parameters $tR$, $\th$ and $\phi$. 

Since all the interactions we are including preserve $\bZ_2$, our Hamiltonian has vanishing matrix elements between states of opposite $\bZ_2$ parity. As a result, the spectrum of the even sector can be found from diagonalising only the block of the Hamiltonian that acts on the $\bZ_2$ even states in the truncated basis. Our numerical results are obtained by diagonalising only this block.

Next, we will  discuss the convergence of the spectrum as a function of the truncation energy $\Delta_T$. 
Then we will discuss the phase diagram of this QFT, and show that depending on the angles $(\theta,\phi)$ we may reach ${\cal PT}$-broken or ${\cal PT}$-symmetric phases in the IR. 
Finally   we will interpret the spectrum, at a critical point on the boundary separating the two phases, in terms of the IR CFT $M(3,8)$.

\subsection{Convergence of the spectrum in  $\Delta_T$}

\begin{figure}[t]
	\begin{center}
		\includegraphics[width=0.4\columnwidth]{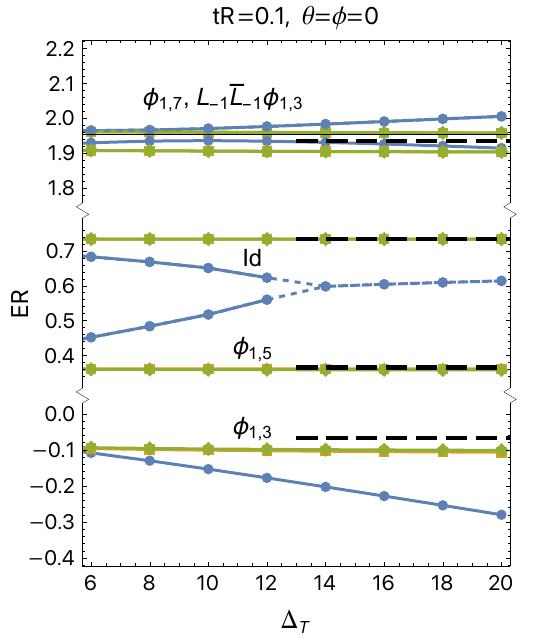}
	\end{center}
	\caption{Convergence of the spectrum at weak coupling. Black dashed lines correspond to first order perturbation theory.  For the maximum value of the cutoff $\D=20$, the Hilbert space (in the $\bZ_2$ even sector) contains 5914 states.}
	\label{figconv}
\end{figure}

Figure~\ref{figconv} illustrates the convergence of the spectrum for a value of the coupling that is weak, where perturbation theory is reliable. 
In blue the spectrum obtained by diagonalising $H_\text{Bare}$, in orange the spectrum  of $H_\text{Bare} +   K_{\text{eff},2}$, and  in green the spectrum of the renormalized Hamiltonian \reef{todiag}. 

Only the green is expected to converge as $\Delta_T\rightarrow\infty$. However, for    these value of the couplings, the spectrum of the second (orange) and third  (green) order Hamiltonian are indistinguishable: both show show a flat, converged, spectrum as a function of $\Delta_T$.

We show first order perturbation theory results with dashed black lines. For our choice of the renormalization scheme, the second and third order corrections to the identity state energy are set to zero, thus the `Id'-state energy is expected to lie almost on top of the black line. For the other states we expect small but visible deviations due to the perturbative corrections beyond first order. The two highest energy states are degenerate to first order in perturbation theory, for $(\th,\phi)=(0,0)$. 

A sensitivity to ultraviolet (UV) divergences is apparent in Figure~\ref{figconv}, as highlighted by the behaviour of the second and third blue lines, which become imaginary for $\Delta_T>12$.  For energy levels that are complex, we plot only the real part using a coloured, dashed line. On the other hand, the renormalised lines (depicted in green and orange) remain real and exhibit a flat, converged behaviour. Since perturbation theory only gives real values for energy levels, the emergence of imaginary parts in the weakly coupled spectrum of the Bare Hamiltonian illustrates its unsuitability as an approximation to the underlying field theory.

\begin{figure}[t]
	\begin{center}
		\includegraphics[width=0.3\columnwidth]{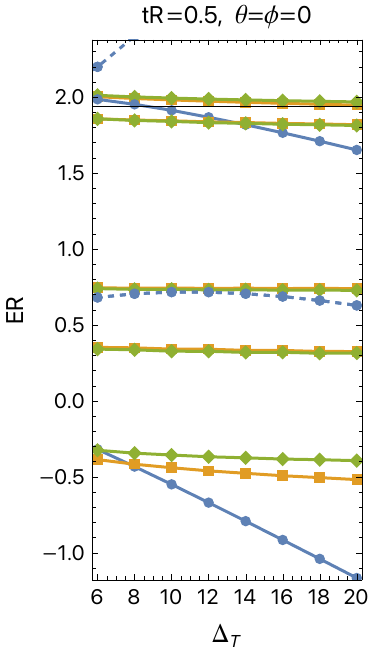}
		\includegraphics[width=0.3\columnwidth]{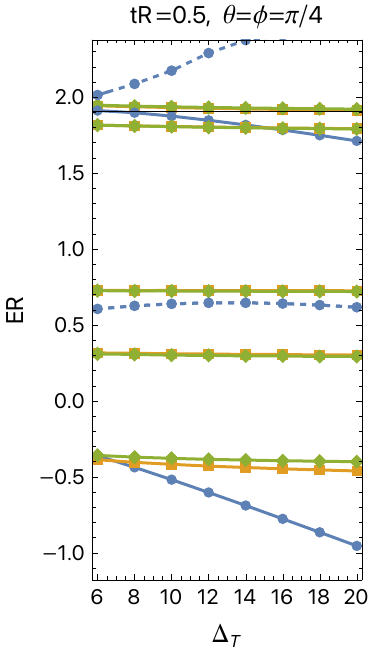}
		\includegraphics[width=0.3\columnwidth]{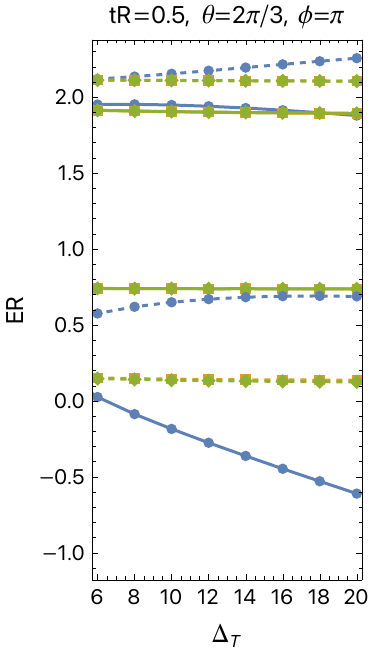}
	\end{center}
	\caption{Convergence of the spectrum at moderate coupling.}
	\label{figconvtwo}
\end{figure}

Having established good agreement between the spectrum of the truncated Hamiltonian \reef{todiag}  and the perturbative regime, we now increase the coupling strength. The results are presented in Figure~\ref{figconvtwo}. Once again, we observe that the bare HT approach produces an unconverged spectrum as $\Delta_T$	
  increases, whereas the renormalized results (depicted in green) demonstrate much flatter and convergent behaviour.
The figure shows  three example angles; however, convergence has been observed across the entire sphere for this value of the coupling. Notably the point $(0,0)$, at which only the renormalized $g_7$ coupling is nonzero, represents one of the least favourable cases for convergence. This highlights the robustness of the renormalization scheme in the moderately strongly coupled regime.

Differences between the orange line (representing the Hamiltonian with only second-order corrections) and the green line (representing the fully renormalized Hamiltonian) are now more pronounced. At the scales of $\Delta_T$   shown in the plot, the overall differences between the orange and green lines remain modest. However, we observe for the lowest energy state that the variation in the green line is noticeably smaller than that for the orange. This behaviour is consistent with expectations, indicating that $H_\text{bare}+K_\text{eff,2}+K_\text{eff,3}$ exhibits better convergence properties than $H_\text{bare}+K_\text{eff,2}$, and we expect the latter to diverge. 
 This result reinforces the reliability of the renormalized approach even under stronger coupling conditions.

 On the first two plots, we observe the same phenomena we saw in the perturbative regime: the presence of complex levels in the spectrum of $H_\text{bare}$ (indicated by dashed blue lines) which become real once renormalization corrections are accounted for. On the other hand, the third plot shows a qualitatively different result. For this choice of couplings, the lowest eigenvalue of $H_\text{bare}$ is real and badly converged (see lowest blue line), whereas the renormalized Hamiltonian shows a genuine well converged but complex lowest eigenvalue. This is due to a physical 'spontaneous breaking' of ${\cal PT}$ symmetry. As explained above, ${\cal PT}$-symmetry ensures that the Hamiltonian's spectrum is either entirely real or contains complex eigenvalues that appear in conjugate pairs. We interpret these complex levels in the spectrum as signs that ${\cal PT}$-symmetry is spontaneously broken. 
  
  Having established that, for the moderately strong coupling $tR=0.5$, the HT calculations remain well-controlled and exhibit good convergence, we now turn our attention to a discussion of the ${\cal PT}$ symmetry phases.

\subsection{Phase Diagram}

\begin{figure}[t]
	\begin{center}
		\includegraphics[height=0.3\columnwidth, trim=0.6cm 1.5cm 0.6cm 0.5cm, clip]{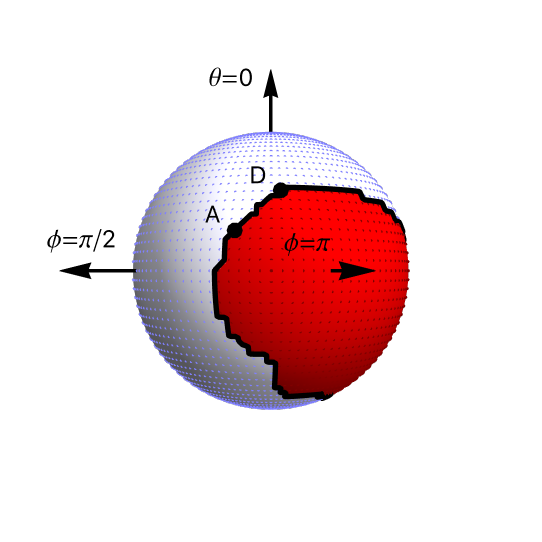}
		\includegraphics[height=0.3\columnwidth, trim=0.2cm 1.5cm 0.6cm 0.5cm, clip]{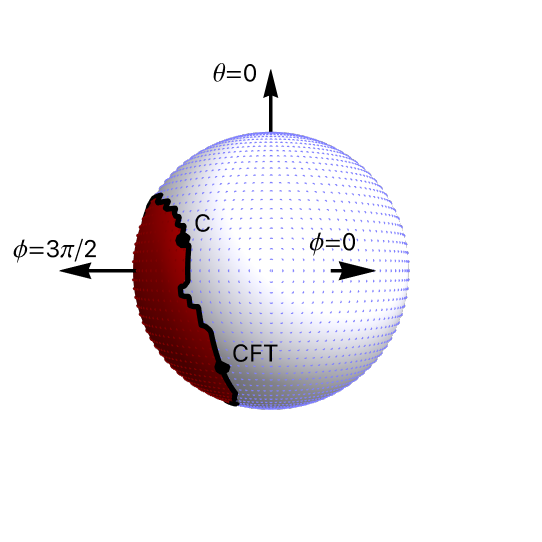}\quad
		\includegraphics[height=0.27\columnwidth]{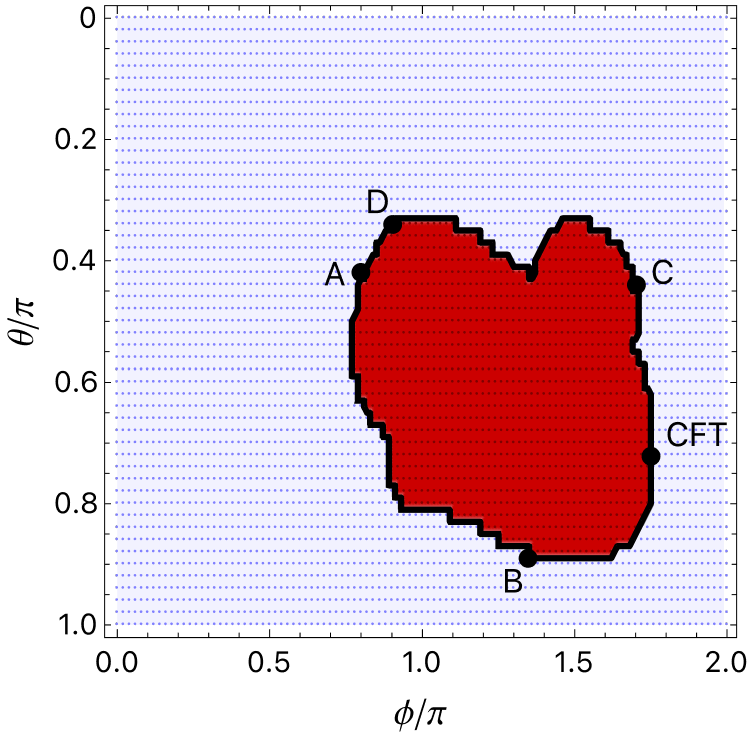}
	\end{center}
	\caption{Phase diagram of the deformed $M(3,10)$ CFT. ${\cal PT}$ symmetry is broken in the dark red shaded region. Data is taken for a radius $tR=0.5$ and at a truncation level $\D=18$. The plot is produced by calculating the spectrum at each point shown in the figures and shading it red if any of the lowest five energy levels has an imaginary part. The labels $A,\dots , D$ and CFT are explained in the text.  }
	\label{PDfig}
\end{figure}

To determine the phase diagram we use the lowest four energy gaps (i.e. lowest 5 energy states), 
for which convergence in $\Delta_T$ has been demonstrated in Fig.~\ref{figconvtwo} of the preceding section. When any of the four-energy gaps becomes complex, we label the theory with those couplings as being inside the ${\cal PT}$-broken phase. Conversely, when all the lowest four-energy gaps are real we cannot be certain that we're in the ${\cal PT}$-symmetric phase because we do not have access to the entire field theory spectrum and cannot exclude the possibility of a higher energy state having complex eigenvalue. 

In Fig.~\ref{PDfig} we show the phase diagram, at a fixed value of the radius $tR=0.5$. In the $tR\rightarrow\infty$ limit, we may expect to find an IR CFT at the boundary between the phases. At moderate $tR$, we may expect the spectrum to be close to that of the CFT, and well described by a CFT deformed with irrelevant operators. 

Because of the renormalisation scheme being used, the direction $(g_3,g_5,g_7)=(0,0,\pm t^{4/5})$ is not expected to correspond to the RG-flow  $M(3,10)+\phi_{1,7}\rightarrow M(3,8)$. Instead it is expected to lie some distance away from $g_3,g_5=0$. This issue is analysed in detail in the next section. 

In passing we comment on the phase diagram in the $g_7=0$ plane. The direction $(\th,\phi)=(\pi/2,\pi)$ corresponds to perturbing $M(3,10)$ by `$-g_3\phi_{1,3}$' only.  According to Fig.~\ref{PDfig}, this direction is well inside the ${\cal PT}$-broken phase. This observation is compatible with the semi-classical analysis in \cite{Katsevich:2024sov}. There it is shown that the `$+g_3\phi_{1,3}$' perturbation flows to a massive phase, with kinks and breathers. Nevertheless the semiclassical analysis also reveals that the `$-g_3\phi_{1,3}$' direction corresponds to an unstable potential, compatible with our findings. 

Ref.~\cite{Kausch:1996vq} studied the RG-flow $M(3,10)+\phi_{1,5}$ via the direct product formulation $M(3,10)_D+\phi_{1,5}=(M(2,5)+\phi)\otimes (M(2,5)+\phi)$. This RG-flow has also been recently computed in Ref.~\cite{Katsevich:2024sov}. This corresponds to the direction  $(\th,\phi)=(\pi/2,\pi/2)$ in our plot, i.e.`$+\phi_{1,5}$'. Our results do not show  ${\cal PT}$-symmetry breaking in this direction and are compatible with their findings. Instead, the direction `$-\phi_{1,5}$', i.e. $(\th,\pi)=(\pi/2,3\pi/2)$, shows ${\cal PT}$-symmetry breaking, consistent with the spectrum of the  $M(2,5){-}\phi$  theory. Next we investigate the spectrum along the boundary of the ${\cal PT}$-symmetry breaking phase.

\subsection{$M(3,8)$-EFT fits}

\begin{figure}[t]
	\begin{center}
		\includegraphics[height=0.25\columnwidth]{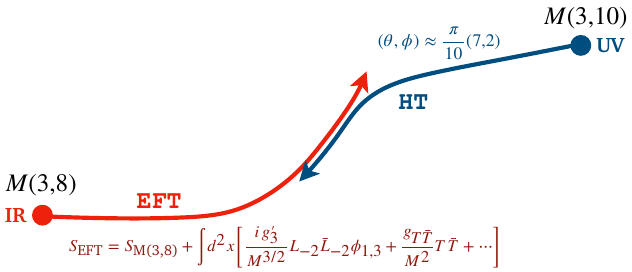}
	\end{center}
	\caption{Schematic representation of the RG-flow studied in this work. The strongly coupled regime of  $M(3,10)+\phi_{1,7}$  can be described by the $M(3,8)$  perturbed by  irrelevant operators. }
	\label{Flowfig}
\end{figure}

Ref.~\cite{Klebanov:2022syt} argued that if the theory $M(3,10)+\phi_{1,7}$  flows to  a CFT in the IR, then the minimal model   $M(3,8)$ is the only possible candidate for such IR CFT.
A further non-trivial check in favour of this proposal, based on preservation  of non-invertible lines along the flow, was given in Ref.~\cite{Nakayama:2024msv}. Table~\ref{tab:M38-evenops} summarizes the CFT data of the even sector of the $M(3,8)$. 

In this section we compare the spectrum along the critical boundary shown in Fig.~\ref{PDfig}, with that of the $M(3,8)$ EFT. This EFT is defined by perturbing the IR $M(3,8)$ CFT by irrelevant operators. In Fig.~\ref{Flowfig} we summarise the logic motivating the comparison presented in this section. The HT calculation allows us to reliably compute the spectrum up to $t R\approx 0.5$. Here we expect that HT at strong coupling coincides with the EFT description, built around the IR fixed point.  As indicated by the red and blue arrows, we will find that there is a regime for which these two calculations overlap.  

Besides the identity, there are three scalar primaries, two of which are relevant and $\mathcal{PT}$ odd and one of which is irrelevant and $\mathcal{PT}$ even. We will  fit the spectrum obtained from our HT computation with an EFT of the form $S_{M(3,8)}+\delta S$, with $\delta S$ containing only scalar, $\mathcal{PT}$-even operators. 
We will need the following fusion rules of the   $M(3,8)$  CFT:
\be\phi_{1,3}\times \phi_{1,3} \sim \mathbb{1} + i\phi_{1,3}+i\phi_{1,5}, \quad \phi_{1,5}\times\phi_{1,5}\sim \mathbb{1}+i\phi_{1,3}+i\phi_{1,5}\,.\ee
We have indicated with an $i$ the OPE coefficients which were purely imaginary.

The fusion rule structure implies that, at first order in perturbation theory, the primary $\phi_{1,7}$ does not correct states in the $\phi_{1,3}$ or $\phi_{1,5}$ conformal family. 
Therefore, even though the $\phi_{1,7}$ is the least irrelevant operator, we can neglect it in our analysis.

\begin{table}[t]
    \centering
    \begin{tabular}{l l l l l l}
 $M(3,8)$ $\mathbb{Z}_2$-even ops.       \phantom{--} & $\phi_{1,1}$   \phantom{--} & $\phi_{1,3}$    \phantom{--} & $\phi_{1,5}$   \phantom{--} & $\phi_{1,7}$    \phantom{\Big|} \\
\hline
\hline
    $\Delta$ &  0  & -1/2& 1/2 & 3  \phantom{\Big|}\\
\hline
$\mathcal{PT}$ & Even & Odd & Odd & Even   \phantom{\Big|} \\
\hline
    \end{tabular}
    \caption{The  $\mathbb{Z}_2$-even scalar primaries of the $M(3,8)$  model,   their dimensions and ${\cal PT}$-symmetry.}
    \label{tab:M38-evenops}
\end{table}

Apart from the two primaries $\phi_{1,3}, \phi_{1,5}$, the next most relevant scalar operator (that is not a total derivative) that we could add to our EFT is the $L_{-2}\bar{L}_{-2}\phi_{1,3}$ descendant, of scaling dimension $4+\Delta_{1,3}=3.5$. We will also take into account the $L_{-2}\bar{L}_{-2}\mathbb{1} = T\bar{T}$ descendant of dimension 4, yielding an EFT action of the form $S=S_{M(3,8)}+\delta  S$, with:
\be
\delta S=\int d^2x \left[\Lambda^2 +\frac{ig_3'}{2\pi M^{3/2}} (L_{-2}\bar{L}_{-2}\phi_{1,3})(x)+\frac{g_{T\bar{T}}}{2\pi  M^2}T\bar{T}(x)\right]\,.
\label{m38eft}
\ee
Here we have denoted the cosmological constant by $\Lambda$, and $M$ the dimensionful cutoff of our EFT, while the $g$'s are dimensionless Wilson coefficients.

We can now perform standard first order perturbation theory with the  action~\reef{m38eft}, computing the corrections to the energy gaps $G_i\equiv (E_i-E_\mathbb{1})$ defined on the cylinder.~\footnote{Going beyond first order would require introducing new operators and couplings constants in \reef{m38eft}. } Below we give the first order result for the four lowest $\mathbb{Z}_2$-even gaps in the $M(3,8)$:
 \begin{align}
RG_{\phi_{1,3}} =& \Delta_{1,3}+\frac{g_{T\bar{T}}}{4(MR)^2}\left[\left(\Delta_{1,3}-\frac{c}{12}\right)^2-\left(\frac{c}{12}\right)^2\right]+i \frac{g_3^\prime}{(MR)^{3/2}}C^3_{33}\alpha_{\phi_3}, \label{gone}\\
RG_{\phi_{1,5}} =& \Delta_{1,5}+\frac{g_{T\bar{T}}}{4(MR)^2}\left[\left(\Delta_{1,5}-\frac{c}{12}\right)^2-\left(\frac{c}{12}\right)^2\right]+i\frac{g_3^\prime}{(MR)^{3/2}}C^3_{55}\alpha_{\phi_5},\\
RG_{\partial\bar{\partial}\phi_{1,3}}=&  \Delta_{1,3}+2+\frac{g_{T\bar{T}}}{4(MR)^2}\left[\left(\Delta_{1,3}+2-\frac{c}{12}\right)^2-\left(\frac{c}{12}\right)^2\right]+i \frac{g_3^\prime}{(MR)^{3/2}}C^3_{33}\beta_{\phi_3},
\\
RG_{\partial\bar{\partial}\phi_{1,5}}=&  \Delta_{1,5}+2+\frac{g_{T\bar{T}}}{4(MR)^2}\left[\left(\Delta_{1,5}+2-\frac{c}{12}\right)^2-\left(\frac{c}{12}\right)^2\right]+i \frac{g_3^\prime}{(MR)^{3/2}}C^3_{55}\beta_{\phi_5}\,.   \label{gfour}
\end{align} 
The $\alpha$'s and $\beta$'s are:
\be \label{EFT-coeffs-m38}
\alpha_{\phi_3} = \frac{1}{1024}\approx 0.00098,\quad \alpha_{\phi_5}=\frac{289}{1024}\approx 0.28,\quad \beta_{\phi_3}= \frac{841}{65536} \approx 0.013, \quad \beta_{\phi_5}=\frac{259081}{65536}\approx 3.953\,.
\ee
The full derivation is given in App.  \ref{EFT-computations}. We also give the relevant OPE coefficients:~\footnote{The absolute value of these OPE coefficients can be found e.g. in Ref.~\cite{Esterlis:2016psv}. Here we determined their sign computing $\langle \phi_{1,3} \phi_{1,3} \phi_{1,5} \phi_{1,5}\rangle$ using  the Coulomb gas formalism \cite{Dotsenko:1984nm}.}
\be \label{eq:OPEcoefs-M38}
C^3_{33}= i\frac{\Gamma(\sfrac{1}{8})\sqrt{-\frac{\Gamma(\sfrac{5}{4})}{\Gamma(\sfrac{-1}{4})}}}{2^{3/4}\Gamma(\sfrac{7}{8})}\approx 1.768i\,, \quad C^3_{55}=i\frac{2^{27/8}\sqrt{-\frac{\Gamma(5/8)}{\Gamma(\sfrac{-1}{4})}}\left(\Gamma(\sfrac{9}{8})\right)^{3/2}}{\pi^{1/4}\Gamma(\sfrac{-1}{8})}\approx -0.442i\,.
\ee

\begin{figure}[t]
	\begin{center}
		\includegraphics[width=.7\columnwidth]{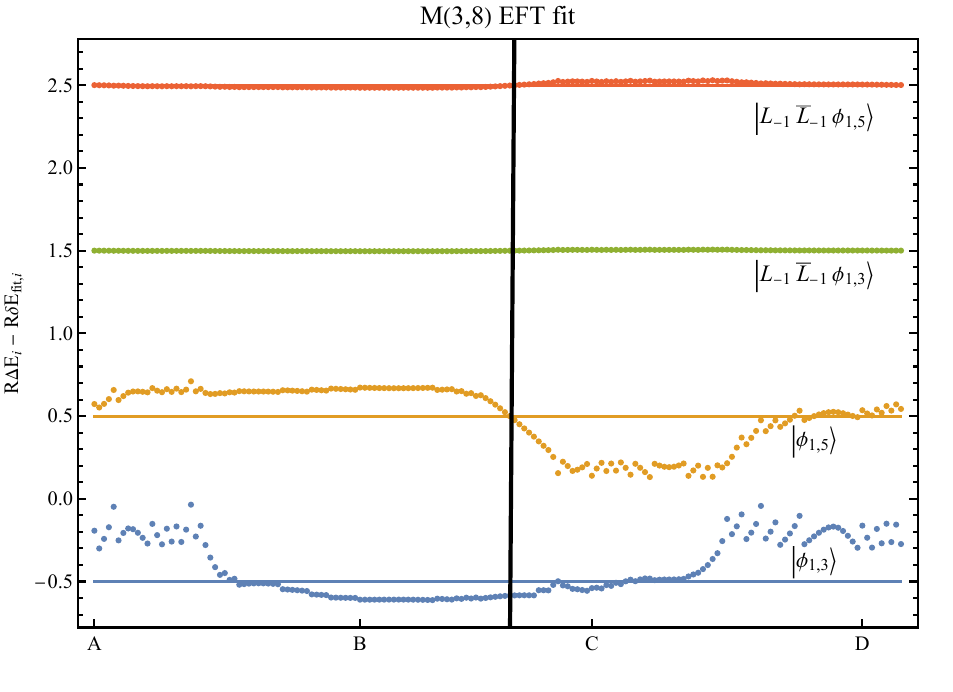}
	\end{center}
	\caption{The black vertical line indicates the point on the boundary where the $\chi^2$ function is smallest, and corresponds to angles of $(\theta,\phi)\approx \pi(0.7,1.8)$. This point is labelled as CFT in Fig. \ref{PDfig}. The labels $A,B,C,D$ refer to the analogous points on the same figure.  }
	\label{fits1}
\end{figure}

In Fig.~\ref{fits1} we show the results of a $\chi^2$-minimization, $\chi^2(g_{T\bar T},g_3^\prime)=R^2\sum_i (G_i-\Delta E_i)^2 $,  where the indices of the sum run over the expressions in \reef{gone}-\reef{gfour} and $\Delta E_i$ is the energy gap calculated using HT. To do the fits, we set $M=1/R$ in the expressions and treat $g_i$ as free parameters.
For each point in the boundary of Fig.~\ref{PDfig} we perform a separate fit. The points along the boundary can be followed through the labels $A,B,C,D$. 
At a particular point between $B$ and $C$ we find the best fit value roughly at a position of $(\theta,\phi)\approx \pi(0.7,0.2) $ (it is indicated with a 'CFT' point in Fig.~\ref{PDfig}). 
The exact location on the $(\theta, \phi)$-plane of the best fit value  depends on the coupling $tR$. At the best fit point, minimising the $\chi^2$ selects values of $g_3'\approx -0.04$ and $g_{T\bar{T}}\approx -0.30$. Plugging in these values in Eqs. \reef{gone}-\reef{gfour}, one can verify that for these couplings the corrections to the CFT energies are perturbative. 

The lowest gap shows a substantial discrepancy for the best fit value (of roughly $10\%$). We attribute this to the fact that this is the worst converged line in Fig.~\ref{figconvtwo}. Extrapolation in $\Delta_T$ of this energy level shows a variation compatible with this discrepancy. 
On the other hand the higher excited states are nailed down for the best fit values. These are also the best converged lines. The fit is of good quality, and  shows evidence supporting the conjecture of Refs.~\cite{Fei:2014xta,Klebanov:2022syt,Katsevich:2024jgq}.

Note that in this work we have only focused on the even sector of $M(3,10)$, for which there is no difference between the A and D-modular invariant theory.  
The theory $M(3,10)+\phi_{1,7}$ is expected to flow to $M(3,8)$ in either the A or D series cases~\cite{Klebanov:2022syt,Nakayama:2024msv}. Fig.~\ref{fits1} is therefore also evidence for the $M(3,10)_A+\phi_{1,7}\rightarrow M(3,8)_A$ version of the flow.~\footnote{We thank I.~Klebanov for emphasising this to us. }

\begin{figure}[h]
	\begin{center}
		\includegraphics[width=.9\columnwidth]{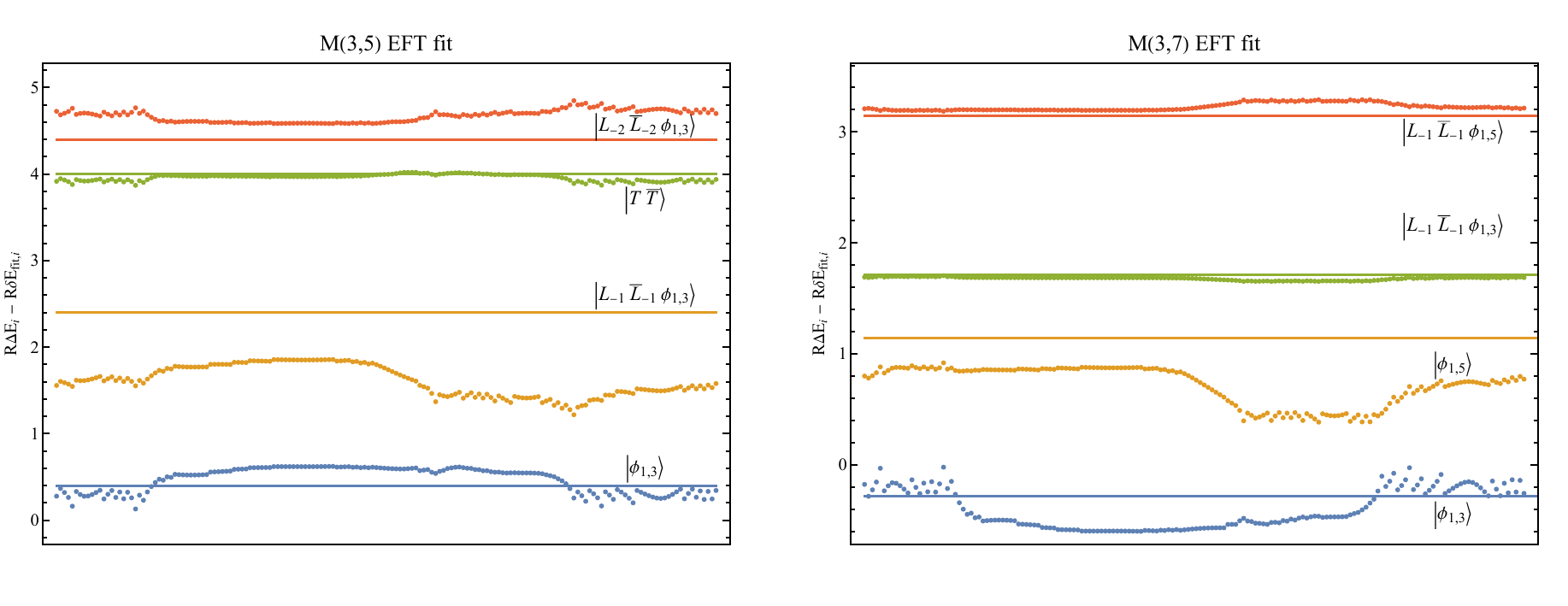}
	\end{center}
	\caption{Fits of the $M(3,5)$ and $M(3,7)$ EFTs. It is apparent from the large disagreement between the orange line and the CFT value, that neither of these fits work.}
	\label{fits2}
\end{figure}

Next we show further  evidence that the fits in Fig.~\ref{fits1} are non-trivial, and that the fitting procedure is able to distinguish between different possible IR CFTs. To that end we have fitted the spectrum with two additional EFTs, derived from the $M(3,5)$ and the $M(3,7)$ CFTs respectively. We take the actions of the two EFTs to be:
\begin{align}
S_{M(3,5)}^{\text{EFT}}=& S_{M(3,5)}+ \int d^2x\left[\Lambda^2 +i  \frac{g_3}{2\pi}M^{8/5}\phi_{1,3}(x) + \frac{g_{T\bar{T}}}{2\pi M^2}T\bar{T}(x)\right]\, \label{action-M35} ,\\
S_{M(3,7)}^{\text{EFT}}=&S_{M(3,7)}+\int d^2x\left[\Lambda^2 +i  \frac{g_3'}{2\pi M^{12/7}}(L_{-2}\bar{L}_{-2}\phi_{1,3})(x) + \frac{g_{T\bar{T}}}{2\pi M^2}T\bar{T}(x)\right]\,.\label{action-M37} 
\end{align}

A flow to the $M(3,5)$ or $M(3,7)$ IR CFT would respect the generalised $c$-theorem~\cite{Castro-Alvaredo:2017udm} 
 because their $c_\text{eff}^\text{IR}=1-6/pq$ are lower than the $c_\text{eff}$ of the $M(3,10)$.
It is important to note that the $M(3,5)$ and $M(3,7)$ have an anomalous $\bZ_2$-symmetry and thus do not have D-series modular invariant.
These CFTs cannot be in the IR of $M(3,10)+\phi_{1,7}$. Nevertheless we will fit our numerical data to these EFTs as a test of our HT method. 

Both fits shown in Fig.~\ref{fits2} are rather poor.
Comparison of the fits in Fig.~\ref{fits1} and Fig.~\ref{fits2} reveals that  \emph{not everything goes}, i.e.  the fitting procedure is a good diagnostic for choosing between various candidate IR CFTs. 
In the future it will be interesting to investigate the scaling with $R$ of all  these different EFT-fits, and also to achieve larger coupling in the UV HT calculation. 

\section{Conclusions}

This work extends the application of Hamiltonian Truncation (HT) methods to study the challenging RG-flow from the \(M(3, 10)\) conformal field theory (CFT) deformed by the relevant primary operator \(\phi_{1,7}\) to the \(M(3, 8)\) CFT. By addressing the technical obstacles associated with the renormalization of the UV theory up to third order in the coupling constant, we successfully demonstrated the efficacy of HT in capturing the non-perturbative spectrum of this flow.

In this work, we have determined the local counterterms for the \(M(3,10) + \phi_{1,7}\) theory. An intriguing future direction would be to identify a renormalization scheme that preserves the topological lines described in~\cite{Nakayama:2024msv}. Achieving this would involve removing the coefficients of the order $\epsilon^{0}$ terms of $\phi_{1,3}$ and $\phi_{15}$ operators from \reef{ct2}, \reef{ct2des} and \reef{eq:ct-3order-phi3}. It would also be interesting to achieve larger values of $\Delta_T$. This would enable more  detailed EFT studies, such as investigating the scaling of the fitted EFT couplings as a function of the  radius. 

The renormalization process, characterised by a combination of counterterms and Effective Hamiltonians, has revealed strong consistency with the $c$-theorem and symmetries of the theory. Our numerical results confirm the convergence of the spectrum even in the strong coupling regime and provide additional evidence supporting the proposed flow 
$
M(3, 10) + \phi_{1,7} \rightarrow M(3, 8)
$.

This study not only demonstrates the versatility of HT methods in two-dimensional QFTs, but also sets the stage for further investigations into complex RG-flows, involving higher-order couplings and more intricate CFTs. Future work could explore extending these methods to higher-dimensional or non-minimal models to deepen our understanding of strongly coupled systems.

\section*{Acknowledgements}

We thank I.~Klebanov for very useful comments on the draft. 
This project has received support from the European Research Council, grant agreement n. 101039756.

\appendix

\section{$M(3,10)_D \cong M(2,5) \otimes M(2,5)$ }
\label{m310}

When a 2D CFT is placed on a   torus, the
requirement of modular invariance   introduces a critical constraint. 
 This condition ties the holomorphic and antiholomorphic components together in a highly non-trivial way. See \cite{DiFrancesco:1997nk} for a pedagogical introduction.

For minimal models $M(p,q)$, modular invariance governs how primary fields from the holomorphic and antiholomorphic sectors are combined, leading to different possible modular invariants. These modular invariants correspond to distinct consistent spectra of the theory on the torus, encoded in modular invariant partition functions. In the following, we focus on the D-series modular invariant of $M(3,10)$, which can be expressed in terms of the tensor product of two copies of the diagonal modular invariant of $M(2,5)$. Notably, $M(2,5)$ has a unique modular invariant, corresponding to a diagonal spectrum.


\subsection*{Partition function of the $M(2,5)$ and $M(3,10)_{D_6}$}

The Hilbert space and the   partition function of the $M(2,5)$ on a torus is given by 
\be \label{eq:Z-M25}
\mathcal{H}_{M(2,5)} = \left(\mathcal{R}_0 \otimes \overbar{\mathcal{R}}_0\right) \oplus \left(\mathcal{R}_{-1/5} \otimes \overbar{\mathcal{R}}_{-1/5}\right), \quad Z_{M(2,5)}(\tau) = |\chi_{1,1}(\tau)|^2 + |\chi_{1,2}(\tau)|^2.
\ee
The notation $\mathcal{R}_h, \overbar{\mathcal{R}}_{\bar{h}}$ describes the irreducible representation of the holomorphic/antiholomorphic Virasoro algebra, respectively (recall that they depend implicitly on the central charge of the theory in question, in this case $c=c_{2,5}=-22/5$).
The spectrum of the theory is composed of two scalar primaries $\mathbb{1}, \phi$ of conformal weights: $h=\bar{h}=h_{1,1} =0$ and $h= \bar{h}=h_{1,2}=-1/5$ respectively. The torus partition function is a function of the modular parameter $\tau$, the ratio of the two periods of the plane lattice defining the torus. The characters of an irreducible representation of the Virasoro algebra are defined as follows:  
\be \chi_{r,s}(\tau) = \text{Tr}_{\mathcal{R}_{h_{r,s}}}[q^{L_0-c/24}], \quad \overbar{\chi}_{r,s}(\bar{\tau}) = \text{Tr}_{\bar{\mathcal{R}}_{h_{r,s}}}\left[\bar{q}^{\overbar{L}_0-c/24}\right],
\ee
where $q=e^{2\pi i \tau}, \bar{q}=e^{-2\pi i \bar{\tau}}$. From the form of the partition function expressed in terms of characters and their complex conjugates as in \eqref{eq:Z-M25}, one can deduce the spectrum of the corresponding theory. The presence of $n$ copies of the term $\chi_{r,s}(\tau)\bar{\chi}_{t,u}(\bar{\tau})$ in the partition function corresponds to $n$ copies of a primary operator with conformal weights $(h_{r,s},h_{t,u})$, scaling dimension $h_{r,s}+h_{t,u}$ and spin $|h_{r,s}-h_{t,u}|$.

The partition function for the $D_6$ modular invariant of the $M(3,10)$ reads:
\be \label{eq:Z-M310}
Z^{M(3,10)}_{D_6}(\tau) = |\chi_{1,1}(\tau)+\chi_{1,9}(\tau)|^2+|\chi_{1,3}(\tau)+\chi_{1,7}(\tau)|^2 +2|\chi_{1,5}(\tau)|^2.
\ee
In the spectrum of this theory we find the five scalar, $\bZ_2$ even primaries of the diagonal modular invariant $M(3,10)$, as well as an additional copy of the $\phi_{1,5}$ field. Furthermore, there are four spinning primaries. Two have both spin and scaling dimension equal to two, while the last two primaries have scaling dimension $1/5$ and spin one. The ten primaries of the $M(3,10)_{D_6}$ are summarized in Table \ref{tab:M310D6}.\footnote{The $\mathcal{PT}$ charges of $\bZ_2$-odd primaries are not fixed by the fusion rules alone: only the relative charge of two operators is unambiguous. However in this case the $\mathcal{PT}$ charge of $\psi_{1,5}$ can be assigned  from the Ginzburg-Landau description \cite{Katsevich:2024jgq}, and therefore the fusion rules \eqref{eq:fusion-rules-odd} fix the rest of the charges.}

\begin{table}[h]
    \centering
    \begin{tabular}{l l l l l l l l l l l }
 $M_{3,10}^{D_6}$ primaries  &  $\phi_{1,1}$ \phantom{--} & $\phi_{1,3}$   \phantom{--}  & $\phi_{1,5}$   \phantom{--} & $\psi_{1,5}$    \phantom{--}  & $\phi_{1,7}$     \phantom{--} & $\phi_{1,9}$   \phantom{--} & $\Phi_{1}$ \phantom{--} & $\Phi_{2}$ \phantom{--} & $\varphi_{1}$ \phantom{--} & $\varphi_{2}$   \phantom{\Big|} \\
\hline
\hline
    $\Delta$ &  0  & -4/5 & -2/5 &-2/5 & 6/5 & 4 & 2 & 2 & 1/5 & 1/5   \phantom{\Big|}\\
\hline
$\bZ_2 \quad (1 \leftrightarrow 2)$ & + & + & + & --  & + & + & -- & -- & -- & -- \phantom{\Big|} \\
\hline
$\mathcal{PT}$ & + & + &  -- & +& + & + & -- &--  & -- & --  \phantom{\Big|} \\
\hline
    \end{tabular}
    \caption{The ten primaries of the $D_6$ modular invariant of the $M(3,10)$ minimal model along with their scaling dimensions and transformation properties under the exchange of the two $M(2,5)$'s and $\mathcal{PT}$ symmetry.}
    \label{tab:M310D6}
\end{table}

\subsection*{$M(2,5)\otimes M(2,5)$ tensor product and the  $D_6$ modular invariant of $M(3,10)$}

\subsubsection*{Hilbert space identification and identities between characters}

Given the Hilbert space structure of the $M(2,5)$, the tensor product $M_{2,5}^{(1)}\otimes M^{(2)}_{2,5}$ decomposes as:
\be \label{eq:H-M25product}
\mathcal{H}_{M_{2,5}^{(1)}\otimes M^{(2)}_{2,5}} = \mathcal{H}_1 \oplus \mathcal{H}_2 \oplus \mathcal{H}_3 \oplus \mathcal{H}_4\, ,
\ee
where 
\begin{align}
\begin{aligned}
\mathcal{H}_1 \equiv& \left(\mathcal{R}_0^{(1)} \otimes \mathcal{R}_0^{(2)}\right) \otimes \left(\overbar{\mathcal{R}}_0^{(1)} \otimes \overbar{\mathcal{R}}_0^{(2)}\right),\quad & \mathcal{H}_2 \equiv &\left(\mathcal{R}_0^{(1)} \otimes \mathcal{R}_{-1/5}^{(2)}\right) \otimes \left(\overbar{\mathcal{R}}_0^{(1)} \otimes \overbar{\mathcal{R}}_{-1/5}^{(2)}\right),\\
\mathcal{H}_3 \equiv&  \left(\mathcal{R}_{-1/5}^{(1)} \otimes \mathcal{R}_0^{(2)}\right) \otimes \left(\overbar{\mathcal{R}}_{-1/5}^{(1)} \otimes \overbar{\mathcal{R}}_0^{(2)}\right),\quad  & \mathcal{H}_4 \equiv&  \left(\mathcal{R}_{-1/5}^{(1)} \otimes \mathcal{R}_{-1/5}^{(2)}\right) \otimes \left(\overbar{\mathcal{R}}_{-1/5}^{(1)} \otimes \overbar{\mathcal{R}}_{-1/5}^{(2)}\right).
\end{aligned}
\end{align}
The claim that $M(3,10)_{A_2,D_6}$ is equivalent to $M(2,5)^{(1)} \otimes M(2,5)^{(2)}$ means that their partition functions must coincide. We now identify precisely how the highest weight states in Table.~\ref{tab:M310D6} and their descendants can be built from states in the tensor product Hilbert space \reef{eq:H-M25product}. 

A quick inspection of the various scaling dimensions reveals that  the primaries of even scaling dimensions, $\phi_{1,1}, \phi_{1,9}, \Phi_1$ and $\Phi_2$ are found in $\mathcal{H}_1$. Likewise, $\mathcal{H}_4$ contains states with scaling dimensions that are sums of scaling dimensions from the $\phi$ family in $M(2,5)$, and hence contains the $M(3,10)$ primaries $\phi_{1,3}, \phi_{1,7}, \varphi_1$ and $\varphi_2$. Finally, $\mathcal{H}_{2}\oplus \mathcal{H}_3$ contains states with scaling dimensions that are sums of scaling dimensions from the $\mathbb{1}$ and $\phi$ families in $M(2,5)$, and hence must contain $\phi_{1,5}$ and $\psi_{1,5}$. 
These statements can be further justified rigorously as we do next. 

We denote the generators of the Virasoro algebras in the first and second copy of $M(2,5)$  by $\{L_n^{(1)}\}, \{\bar{L}^{(1)}_n\}, \{L_n^{(2)}\}$ and $\{\bar{L}_n^{(2)}\}$. Then, the generators of the holomorphic and antiholomorphic Virasoro algebras on the tensor product space are defined to be 
\be \label{eq:tensor-L}
L_n \equiv L_n^{(1)}\otimes \mathbb{1}^{(2)} + \mathbb{1}^{(1)}\otimes L_n^{(2)}, \quad \overbar{L}_n \equiv \overbar{L}_n^{(1)}\otimes \mathbb{1}^{(2)} + \mathbb{1}^{(1)}\otimes \overbar{L}_n^{(2)}, \quad \mathbb{1} \equiv \mathbb{1}^{(1)}\otimes \mathbb{1}^{(2)}.
\ee 
It can be easily checked that the above generators obey the Virasoro algebra with a central charge equal to the sum of the central charges from each of the components of the tensor product. This is in accordance with $c_{3,10} = 2c_{2,5}$. Furthermore, it's easy to see that 
\be
\text{Tr}_{\mathcal{R}^{(1)}_h\otimes \mathcal{R}^{(2)}_{h^\prime}}\left[q^{L_0^{(1)}\otimes \mathbb{1}^{(2)} + \mathbb{1}^{(1)}\otimes L_0^{(2)}-2c_{2,5}/24}\right] = \text{Tr}_{\mathcal{R}^{(1)}_h}\left[q^{L_0^{(1)}-c_{2,5}/24}\right]\text{Tr}_{\mathcal{R}^{(2)}_{h^\prime}}\left[q^{L_0^{(2)}-c_{2,5}/24}\right] \, 
\ee
Therefore, the partition function of the tensor product theory can be written in terms of the characters of a single copy of $M(2,5)$:
\be \label{eq:Z-product}
Z_{M_{2,5}^{(1)}\otimes M^{(2)}_{2,5}}(\tau) = \left|\chi_{1,1}(\tau)^2\right|^2+ 2|\chi_{1,1}(\tau)\chi_{1,2}(\tau)|^2+ \left|\chi_{1,2}(\tau)^2\right|^2.
\ee
The precise equivalence with Eq. \eqref{eq:Z-M310} is obtained with 
\be \label{eq:characters}
\left(\chi_{1,1}^{M_{2,5}}\right)^2 = \chi_{1,1}^{M_{3,10}}+\chi_{1,9}^{M_{3,10}}, \quad \chi_{1,1}^{M_{2,5}}\chi^{M_{2,5}}_{1,2} = \chi_{1,5}^{M_{3,10}}, \quad \left(\chi_{1,2}^{M_{2,5}}\right)^2 = \chi_{1,3}^{M_{3,10}}+\chi_{1,7}^{M_{3,10}}.
\ee
We have supressed the $\tau$ dependence of characters, and the complex conjugate of the above relation holds for the antiholomorphic characters. These equalities can be checked by using the explicit expression for a character $\chi_{r,s}$ in the minimal model $M(p',p)$:
\be \label{eq:character}
\chi_{r,s}^{M_{p',p}}(q) = \frac{q^{-1/24}}{\varphi(q)}\sum_{n \in \bZ}\left[ q^{\frac{(2p p^\prime n + rp - p^\prime s)^2}{4p p^\prime}} - q^{\frac{(2p p^\prime n + rp + p^\prime s)^2}{4p p^\prime}}\right].
\ee
$\varphi(q) = \prod_{k\geq 1}(1-q^k)$ is the Euler function, the reciprocal of the generating function for the integer partitions. 

\subsubsection*{Explicit dictionary between states}

Let us now give the $M(3,10)$ primaries in the tensor product space. The $\phi_{1,1}$ chiral component in the $M(3,10)$ must have zero eigenvalue under $L_0$, and there is trivially only one such state in the tensor product space. Therefore, \be \label{eq:identitystate}
\ket{\mathbb{1}} = \left(\ket{0}^{(1)}\otimes \ket{0}^{(2)}\right) \otimes \left(\overbar{\ket{0}}^{(1)}\otimes \overbar{\ket{0}}^{(2)}\right).
\ee
 From \eqref{eq:identitystate}, applying raising operators $L_{-m},\overbar{L}_{-n}$ defined in \eqref{eq:tensor-L} generates the entire identity family. Now let's identify the $\phi_{1,9}$ primary, which should have an eigenvalue of two under the operator $L_0$. There are two linearly independent states in the tensor product space with such eigenvalue: $L_{-2}^{(1)}\ket{0}^{(1)}\otimes \ket{0}^{(2)}$ and $\ket{0}^{(1)}\otimes L_{-2}^{(2)} \ket{0}^{(2)}$. We can directly identify the linear combination $L_{-2}\ket{\mathbb{1}} =  L_{-2}^{(1)}\ket{0}^{(1)}\otimes \ket{0}^{(2)}+\ket{0}^{(1)}\otimes L_{-2}^{(2)} \ket{0}^{(2)}$, which belongs to the identity Virasoro multiplet. The orthogonal linear combination is proportional to $L_{-2}^{(1)}\ket{0}^{(1)}\otimes \ket{0}^{(2)}-\ket{0}^{(1)}\otimes L_{-2}^{(2)} \ket{0}^{(2)}$ and should correspond to the chiral part of the $\phi_{1,9}$ state. To check that this state indeed behaves as a primary, one can verify that it is annihilated by $L_2$ (it is trivially annihilated by $L_1$ since $[L_1^{(i)},L_{-2}^{(i)}] \propto L_{-1}^{(i)}$ and $L_{-1}^{(i)}\ket{0}^{(i)} = 0, i=1,2$). This implies that it is annihilated by all $L_{n>0}$, as is needed. Furthermore, we can normalize it and include the anti chiral linear combination in order to write:~\footnote{Under exchange of the two $M(2,5)$'s, the chiral and anti chiral components each pick up a sign, but the full state is even, in the same manner as the primary $\phi_{1,9}$ is $\bZ_2$-even in the diagonal theory. The same goes for the norm: the norm of the chiral and anti-chiral components are both minus one, but the full primary state has unit norm.}
\be \label{eq:phi19}
\ket{\phi_{1,9}} = \frac{5}{22}\left[\ket{T}^{(1)}\otimes \ket{0}^{(2)}-\ket{0}^{(1)}\otimes\ket{T}^{(2)}\right]\otimes\left[\ket{\overbar{T}}^{(1)}\otimes \overbar{\ket{0}}^{(2)}-\overbar{\ket{0}}^{(1)}\otimes \ket{\overbar{T}}^{(2)}\right],
\ee
where $\ket{T}^{(i)}\equiv L_{-2}^{(i)}\ket{0}^{(i)}$ for $i=1,2$ and similarly for the antiholomorphic state. 
The two spinning primary states in $\mathcal{H}_1$ are given by: 
\be \label{eq:Phi1}
\ket{\Phi_1} = \sqrt{\frac{5}{22}}\left[\ket{T}^{(1)}\otimes \ket{0}^{(2)}-\ket{0}^{(1)}\otimes \ket{T}^{(2)}\right]\otimes \left[ \overbar{\ket{0}}^{(1)}\otimes \overbar{\ket{0}}^{(2)}\right],
\ee
\be \label{eq:Phi2}
\ket{\Phi_2} =\sqrt{\frac{5}{22}} \left[ \ket{0}^{(1)}\otimes \ket{0}^{(2)}\right]\otimes \left[\ket{\overbar{T}}^{(1)}\otimes \overbar{\ket{0}}^{(2)}-\overbar{\ket{0}}^{(1)}\otimes \ket{\overbar{T}}^{(2)}\right].
\ee
It is straightforward to check that these states are orthogonal to each other and the other primaries given before (as well as their descendants) and that they have correct conformal weights (i.e. eigenvalues under $L_0, \overbar{L}_0$): $(2,0)$ and $(0,2)$ respectively. We note that under exchange of the two $M(2,5)$'s, both $\ket{\Phi_1}$ and $\ket{\Phi_2}$ pick up a minus sign, and are therefore odd under the global $\bZ_2$ of $D_6$ modular invariant of $M(3,10)$. Finally, we note that they have a norm of minus one.

Let's continue the exercise for $\mathcal{H}_4$. Similarly to the identity, the $\phi_{1,3}$ state is simply the tensor product of the primary states:
\be \label{eq:phi13}
\ket{\phi_{1,3}} = \left(\ket{\phi}^{(1)}\otimes \ket{\phi}^{(2)}\right) \otimes \left(\ket{\overbar{\phi}
}^{(1)}\otimes \ket{\overbar{\phi}}^{(2)}\right),
\ee
We have defined $\ket{\phi}^{(i)} \equiv \ket{-1/5}^{(i)}, i=1,2$ for simplicity and analogously for the antiholomorphic state. The state \eqref{eq:phi13} has the correct eigenvalue under $L_0$ and $\overbar{L}_0$, is annihilated by lowering operators, has norm one and is even under exchange of $M(2,5)$'s, as the primary $\phi_{1,3}$ should. There is only one state at the first level in the $\phi_{1,3}$ family whose chiral component is $L_{-1}^{(1)}\ket{\phi}^{(1)} \otimes \ket{\phi}^{(2)} + \ket{\phi}^{(1)} \otimes L_{-1}^{(2)}\ket{\phi}^{(2)}$. The only orthogonal linear combination with the same eigenvalue under $L_0$ therefore corresponds, up to normalization, to the chiral part of the $\phi_{1,7}$ state, which is: 
\be \label{eq:phi17}
\ket{\phi_{1,7}} = \frac{5}{4}\left(L_{-1}^{(1)}\ket{\phi}^{(1)} \otimes \ket{\phi}^{(2)} - \ket{\phi}^{(1)} \otimes L_{-1}^{(2)}\ket{\phi}^{(2)}\right)\otimes \left(\overbar{L}_{-1}^{(1)}\ket{\overbar{\phi}}^{(1)} \otimes \ket{\overbar{\phi}}^{(2)} - \ket{\overbar{\phi}}^{(1)} \otimes \overbar{L}_{-1}^{(2)}\ket{\overbar{\phi}}^{(2)}\right).
\ee
Once again one can check that this state is annihilated by all $L_{n>0},\overbar{L}_{n>0}$ and is even under the exchange of the two $M(2,5)$'s, coinciding with the symmetry properties of $\phi_{1,7}$ in the diagonal modular invariant. The two spinning primaries in $\mathcal{H}_4$ are:
\be \label{eq:varphi1}
\ket{\varphi_1} = \sqrt{\frac{5}{4}}\left(L_{-1}^{(1)}\ket{\phi}^{(1)} \otimes \ket{\phi}^{(2)} - \ket{\phi}^{(1)} \otimes L_{-1}^{(2)}\ket{\phi}^{(2)}\right)\otimes \left(\ket{\overbar{\phi}
}^{(1)}\otimes \ket{\overbar{\phi}}^{(2)}\right),
\ee
\be \label{eq:varphi2}
\ket{\varphi_2}= \sqrt{\frac{5}{4}}\left(\ket{\phi}^{(1)}\otimes \ket{\phi}^{(2)}\right) \otimes \left(\overbar{L}_{-1}^{(1)}\ket{\overbar{\phi}}^{(1)} \otimes \ket{\overbar{\phi}}^{(2)} - \ket{\overbar{\phi}}^{(1)} \otimes \overbar{L}_{-1}^{(2)}\ket{\overbar{\phi}}^{(2)}\right).
\ee
Finally, let us give the form of the last two primaries, two copies of a scalar of scaling dimension $-2/5$. It is clear from the definition of $\mathcal{H}_{2,3}$ that the two spaces get sent to each other under the exchange of the two $M(2,5)$'s, and therefore the primaries $\ket{\phi_{1,5}}, \ket{\psi_{1,5}}$ with definite charges under the global $\bZ_2$ are linear combinations of states in both $\mathcal{H}_{2,3}$. They are:
\be \label{eq:phi15}
\ket{\phi_{1,5}} = \frac{1}{\sqrt{2}}\left[\left(\ket{0}^{(1)}\otimes \ket{\phi}^{(2)}\right)\otimes \left( \overbar{\ket{0}}^{(1)}\otimes \ket{\overbar{\phi}}^{(2)}\right) + \left(\ket{\phi}^{(1)}\otimes \ket{0}^{(2)}\right)\otimes \left( \ket{\overbar{\phi}}^{(1)}\otimes \overbar{\ket{0}}^{(2)}\right)\right],
\ee
\be \label{eq:phi15}
\ket{\psi_{1,5}} = \frac{1}{\sqrt{2}}\left[\left(\ket{0}^{(1)}\otimes \ket{\phi}^{(2)}\right)\otimes \left( \overbar{\ket{0}}^{(1)}\otimes \ket{\overbar{\phi}}^{(2)}\right) - \left(\ket{\phi}^{(1)}\otimes \ket{0}^{(2)}\right)\otimes \left( \ket{\overbar{\phi}}^{(1)}\otimes \overbar{\ket{0}}^{(2)}\right)\right].
\ee

We have now explicitly given the ten primary states of the $D_6$ modular invariant of $M(3,10)$ in terms of states in the tensor product $M_{2,5}^{(1)}\otimes M_{2,5}^{(2)}$. Half of them are $\bZ_2$ even and half are $\bZ_2$ odd, with the action of $\bZ_2$ being defined as the exchange of the two $M_{2,5}$ constituents of the tensor product. 

\section{CFT data of the $M(3,10)_{D_6}$ from  the tensor product structure}

We will use the tensor product structure $M(3,10)_{D_6} \simeq M(2,5) \otimes M(2,5)$ to compute OPE coefficients.

\subsection*{Fusion rules \& OPE coefficients of the $D_6$ modular invariant}

Next we give the  fusion rules of the ten primaries of the $D_6$ modular invariant. They follow from the fusion rules of the primaries in the diagonal modular invariant, which can be found in e.g. \cite{Klebanov:2022syt}. In order for a primary $\phi_{h_3,\bar{h}_3}$ to appear in the OPE of $\phi_{h_1,\bar{h}_1} \times \phi_{h_1,\bar{h}_1}$, the corresponding chiral and antichiral components must appear in the relevant OPE, i.e : $\phi_{h_3} \in \phi_{h_1} \times \phi_{h_2}, \phi_{\bar{h}_3} \in \phi_{\bar{h}_1} \times \phi_{\bar{h}_2}$. This yields the following fusion rules for the $D_6$ modular invariant: 

\begin{align}
\begin{aligned}\label{eq:fusion-rules-odd}
\psi_{1,5}\times \psi_{1,5}& \sim &\mathbb{1}+ \phi_{1,3} + i \phi_{1,5} + \phi_{1,7}+\phi_{1,9},& \quad & \psi_{1,5} \times \Phi_1 \sim&  \phi_{1,5}, \\
\psi_{1,5} \times \Phi_2 &\sim& \phi_{1,5},& \quad& 
\psi_{1,5} \times \varphi_1 \sim& i\phi_{1,3} + \phi_{1,5} + i\phi_{1,7},\\
\psi_{1,5}\times \varphi_2& \sim&  i\phi_{1,3} + \phi_{1,5} + i\phi_{1,7}, &\quad & 
\Phi_1 \times \Phi_1 \sim &\mathbb{1}+\phi_{1,3} + i \phi_{1,5},\\
\Phi_1 \times \Phi_2& \sim& i\phi_{1,5} + \phi_{1,7}+\phi_{1,9},& \quad &
\Phi_1 \times \varphi_1 \sim& \phi_{1,3},
\\
\Phi_1 \times \varphi_2& \sim &\phi_{1,7},& \quad & 
\Phi_2 \times \Phi_2 \sim &\mathbb{1}+\phi_{1,3} + i \phi_{1,5}, \\
\Phi_2 \times \varphi_1 &\sim &\phi_{1,7}, &\quad & 
\Phi_2 \times \varphi_2 \sim& \phi_{1,3}\, .
\end{aligned}
\end{align}
We have only given the fusion rules of the form $\text{Odd} \times \text{Odd} \sim \text{Even}$.
The $\text{Even} \times \text{Even} \sim \text{Even}$ fusion rules are the same as the diagonal modular invariant, they are given in section~\ref{reviewHT}.

Next we will compute OPE coefficients from the knowledge of the tensor product structure of the $D_6$ invariant, since the only non trivial OPE coefficient in the $M(2,5)$ is known:
\[ C^\phi_{\phi \phi} = \frac{i}{5}\left(\frac{\Gamma(1/5)}{\Gamma(4/5)}\right)^{3/2}\left(\frac{\Gamma(2/5)}{\Gamma(3/5)}\right)^{1/2}.\]
In what follows we will denote $\phi, \overbar{\phi}$ the holomorphic, respectively antiholomorphic component of the non trivial primary in the $M(2,5)$. The non-trivial OPE coefficient in the $M(2,5)$ is related to the three point function of the $\phi \otimes \overbar\phi$ primary: 
\[ C^\phi_{\phi\phi} = \langle \phi(\infty)\phi(1)\phi(0)\rangle \langle \overbar{\phi}(\infty)\overbar{\phi}(1)\overbar{\phi}(0)\rangle. \]

Next,  since $\phi_{1,3} = \phi^{(1)} \otimes \phi^{(2)}, \overbar{\phi}_{1,3} = \overbar{\phi}^{(1)} \otimes \overbar{\phi}^{(2)}$, we have that 
\[ C^3_{33} =  \left(\langle \phi(\infty)\phi(1)\phi(0)\rangle\right)^2 \left(\langle \overbar{\phi}(\infty)\overbar{\phi}(1)\overbar{\phi}(0)\rangle\right)^2 = \left(C^{\phi}_{\phi\phi}\right)^2. \]
Using the tensor product representation of the primary states given in Eqs. \eqref{eq:phi19} \eqref{eq:phi13}, \eqref{eq:phi17} and \eqref{eq:phi15} as well as the operator map correspondance, we can compute all OPE coefficients in terms of known three point functions in the $M(2,5)$. We obtain:
\begin{align}
\begin{aligned}
C^3_{55} &= 1,\quad C^5_{33} = \sqrt{2}C^\phi_{\phi \phi},\quad C^3_{57} = \frac{C^{\phi}_{\phi\phi}}{10\sqrt{2}}\,, \\
C^3_{77} & = \frac{\left(C^\phi_{\phi\phi}\right)^2}{4} \ , \quad 
C^3_{79} = \frac{2}{11} \ , \quad 
C^5_{77} = \frac{49\sqrt{2}}{100}C^\phi_{\phi\phi} \ , \quad   \\
C^5_{55}&= \frac{C^\phi_{\phi\phi}}{\sqrt{2}} \ , \quad 
C^5_{57} = \frac{1}{5} \ , \quad 
 C^5_{59} = \frac{1}{110} \,. 
\end{aligned}
\end{align}

We notice that all OPE coefficients containing an odd number of the primary $\phi_{1,5}$ are proportional to $C^\phi_{\phi\phi}$, and hence they change sign under the change $\Phi^{M_{2,5}} \to -\Phi^{M_{2,5}}$. This can of course be reabsorbed by sending $\phi_{1,5} \to - \phi_{1,5}$, and ultimately any choice of sign for $C^\phi_{\phi\phi}$ yields a consistent definition of the $M(3,10)$ CFT with $D_6$ modular invariant. 

\subsection*{Four-point functions computed from the tensor product structure}
\label{4ptDP}

Now we will compute the needed four-point functions of the $M(3,10)$ using the tensor product structure. 
For this we will  only need the  four-point function of primaries in the $M(2,5)$, which can be found for example in \cite{Mussardo:2010mgq}, and is given by
\be \label{eq:4pt-M25}
\langle \phi(z_1,\overbar{z}_1)\phi(z_2,\overbar{z}_2)\phi(z_3,\overbar{z}_3)\phi(z_4,\overbar{z}_4) \rangle = \left| \frac{z_{13}z_{24}}{z_{12}z_{23}z_{34}z_{14}}\right|^{-4/5}\left\{|F_1(\eta)|^2 + \left(C^\phi_{\phi\phi}\right)^2 |F_2(\eta)|^2\right\}\,,
\ee
where $F_1(\eta) \equiv \, _2F_1(3/5,4/5,6/5;\eta), F_2(\eta) \equiv \eta^{-1/5}\, _2F_1(3/5,2/5,4/5;\eta)$ and $\eta = z_{12}z_{34}/(z_{13}z_{24})$ is the conformally invariant cross ratio.

\subsubsection*{$\langle \phi_{1,3} \phi_{1,7} \phi_{1,7} \phi_{1,7}\rangle$}

In order to compute $\langle \phi_{1,3}(z_1,\overbar{z}_1) \phi_{1,7}(z_2,\overbar{z}_2) \phi_{1,7}(z_3,\overbar{z}_3) \phi_{1,7}(z_4,\overbar{z}_4)\rangle$, we make use of  \eqref{eq:phi13} and \eqref{eq:phi17}. In particular, using state operator correspondance and replacing $L_{-1} \to \partial, \overbar{L}_{-1} \to \overbar\partial$, we have:
\bea
 \label{eq:phi13-op}
\phi_{1,3}(z,\overbar{z}) &= \Phi^{(1)}(z,\overbar{z})\Phi^{(2)}(z,\overbar{z}) \, \\
 \label{eq:phi17-op}
\phi_{1,7}(z,\overbar{z}) &= \frac{5}{4}\left[\partial\overbar\partial \Phi^{(1)}(z,\overbar{z})\Phi^{(2)}(z,\overbar{z}) + \Phi^{(1)}(z,\overbar{z})\partial\overbar\partial \Phi^{(2)}(z,\overbar{z})\right. \nonumber\\
&\left. - \partial \Phi^{(1)}(z,\overbar{z}) \overbar\partial \Phi^{(2)}(z,\overbar{z})- \overbar{\partial} \Phi^{(1)}(z,\overbar{z}) \partial \Phi^{(2)}(z,\overbar{z})\right] \, .
\eea
where $\Phi^{(i)}$ is the field of one of the Lee-Yang copies. 
 Therefore, the four point function we are interested in  can be   organized in eight terms, whose expressions are:
\begin{align} 
\begin{aligned}
\label{eq:groups3777}
\text{I} = &2\left[ (\partial_2\overbar\partial_2\partial_3\overbar\partial_3\partial_4\overbar\partial_4 F)F +(\partial_2\overbar\partial_2\partial_3\overbar\partial_3F)(\partial_4\overbar\partial_4F)\right.\\
 +&\left. (\partial_2\overbar\partial_2\partial_4\overbar\partial_4F)(\partial_3\overbar\partial_3F)+(\partial_2\overbar\partial_2F)(\partial_3\overbar\partial_3\partial_4\overbar\partial_4F)\right],\\
\text{II} =-&2\left[(\partial_2\overbar\partial_2\partial_3\overbar\partial_3\partial_4 F)(\overbar\partial_4 F)+(\partial_2\overbar\partial_2\partial_4F)(\partial_3\overbar\partial_3\overbar\partial_4 F) + \partial \leftrightarrow \overbar\partial \right], \\
\text{III} = -&2\left[(\partial_2\overbar\partial_2\partial_3\partial_4\overbar\partial_4F)(\overbar\partial_3F)+(\partial_3\partial_4\overbar\partial_4F)(\partial_2\overbar\partial_2\overbar\partial_3F)+\partial\leftrightarrow \overbar\partial\right],\\
\text{IV} =&2\left[(\partial_2\overbar\partial_2\partial_3\partial_4F)(\overbar\partial_3\overbar\partial_4F)+(\partial_2\overbar\partial_2\overbar\partial_3\partial_4F)(\partial_3\overbar\partial_4F) + \partial \leftrightarrow \overbar\partial \right], \\
\text{V} =-&2\left[(\partial_2\partial_3\overbar\partial_3\partial_4\overbar\partial_4 F)(\overbar\partial_2F)+(\partial_2\partial_3\overbar\partial_3F)(\partial_4\overbar\partial_4 \overbar\partial_2 F) + \partial \leftrightarrow \overbar\partial\right],\\
\text{VI} =&2\left[(\partial_2\partial_3\overbar\partial_3\partial_4F)(\overbar\partial_2\overbar\partial_4F)+(\overbar\partial_2\partial_3\overbar\partial_3\partial_4F)(\partial_2\overbar\partial_4F)+\partial \leftrightarrow \overbar\partial\right],\\
\text{VII}=&2\left[(\partial_2\partial_3\partial_4\overbar\partial_4F)(\overbar\partial_2\overbar\partial_3F)+(\partial_2\overbar\partial_3\partial_4\overbar\partial_4F)(\overbar\partial_2\partial_3F)+ \partial \leftrightarrow \overbar\partial\right],\\
\text{VIII}=-&2\left[(\partial_2\partial_3\partial_4F)(\overbar\partial_2\overbar\partial_3\overbar\partial_4F)+(\overbar\partial_2\partial_3\partial_4F)(\partial_2\overbar\partial_2\overbar\partial_4F)\right. \\
+&\left.(\partial_2\overbar\partial_3\partial_4F)(\overbar\partial_2\partial_3\overbar\partial_4F)+(\partial_2\partial_3\overbar\partial_4F)(\overbar\partial_2\overbar\partial_3\partial_4F)\right] \, ,
\end{aligned}
\end{align}
where we use the shorthand $F \equiv \langle \phi(z_1,\overbar{z}_1)\phi(z_2,\overbar{z}_2)\phi(z_3,\overbar{z}_3)\phi(z_4,\overbar{z}_4) \rangle$, supressing arguments, and $\partial_i \equiv \partial_{z_i}, \overbar\partial_i \equiv \partial_{\overbar{z}_i}$. The final four-point function is given by 
\be
 \langle \phi_{1,3}(z_1,\overbar{z}_1) \phi_{1,7}(z_2,\overbar{z}_2) \phi_{1,7}(z_3,\overbar{z}_3) \phi_{1,7}(z_4,\overbar{z}_4)\rangle = \left(\frac{5}{4}\right)^3 \left[ \text{I}+\text{II}+\text{III}+\text{IV}+\text{V}+\text{VI}+\text{VII}+\text{VIII}\right] \, . \label{4pt13}
 \ee

\subsubsection*{$\langle \phi_{1,5} \phi_{1,7} \phi_{1,7} \phi_{1,7}\rangle$}

The $\phi_{1,5}$ field is given by
\be
\phi_{1,5}(z,\bar{z}) = \frac{1}{\sqrt{2}} \left( \Phi^{(1)}(z,\bar z)+ \Phi^{(2)} (z,\bar z)\right)\,.
\ee
We use this representation to compute the four-point function $\langle \phi_{1,5} \phi_{1,7} \phi_{1,7} \phi_{1,7}\rangle$.
The expression  is complicated and not practical for large order expansions  in the cross-ratio. Nevertheless we have used this expression   to cross-check alternative calculations   in sections~\ref{CGapp} and \ref{BPZapp}, where we computed four-point functions using the Coulomb Gas and BPZ formalisms, respectively.

\section{Four-point functions via the Coulomb Gas}
\label{CGapp}

In this section we will compute the  two four-point functions via the Coulomb gas method developed by Dotsenko and Fateev~\cite{Dotsenko:1984nm} -- we will follow the notation there explained. 

Starting with $\langle \phi_{1,3} \phi_{1,7} \phi_{1,7} \phi_{1,7}\rangle$,  the chiral conformal block can be obtained by the use of only one screening charge, i.e. 
\be
\langle \phi_{1,3}(z_1)\phi_{1,7}(z_2)\phi_{1,7}(z_3)\phi_{1,7}(z_4)\rangle \sim \langle V_{\alpha_{1,3}}(z_1)V_{\alpha_{-1,-7}}(z_2)V_{\alpha_{-1,-7}}(z_3)V_{\alpha_{-1,-7}}(z_4)Q_+\rangle \, . 
\ee
The integral representation of this Coulomb gas correlator for an arbitrary non trivial contour satisfies the $_2F_1$ ODE.
Imposing the absence of nontrivial monodromies on the full correlator correctly selects the solution to the ODE that corresponds  to the $\phi_{1,5}$ block:
\be \label{eq:4ptfc-phi13}
 \langle\phi_{1,3}(\infty)\phi_{1,7}(1)\phi_{1,7}(z_2)\phi_{1,7}(z_1)\rangle =C_{577}C_{357} \frac{\left|1+z_1^2+z_2^2-z_1-z_2-z_1z_2\right|^2}{|1-z_1|^{14/5}|1-z_2|^{14/5}|z_1-z_2|^{14/5}}\,.
\ee

Next,  we describe the calculation of $\langle \phi_{1,5} \phi_{1,7} \phi_{1,7} \phi_{1,7}\rangle$. The Coulomb Gas correlator with the least amount of screening charges reads:
\[  \langle \phi_{1,5}(z_1)\phi_{1,7}(z_2)\phi_{1,7}(z_3)\phi_{1,7}(z_4)\rangle \sim \langle V_{\alpha_{1,5}}(z_1)V_{\alpha_{-1,-7}}(z_2)V_{\alpha_{-1,-7}}(z_3)V_{\alpha_{-1,-7}}(z_4)Q_+Q_-\rangle \,.\]
The above correlator can be written as a double integral (since there are two screening charges) leaving the contours of integration as of yet unspecified.
Using global conformal invariance to place the field $\phi_{1,5}$ at infinity and two other fields at $0$ and $1$:
\bea
  \langle V_{\alpha_{5,+}}(\infty) &V_{\alpha_{7,-}}(1)V_{\alpha_{7,-}}(z)V_{\alpha_{7,-}}(0)Q_+Q_-\rangle   =z^{4/15}(1-z)^{4/15} \int_{\mathcal{C}_1}dt_1\int_{\mathcal{C}_2}dt_2\\ 
 & \times t_1^{2/5}(t_1-1)^{2/5}(t_1-z)^{2/5} t_2^{-4/3}(t_2-1)^{-4/3}(t_2-z)^{-4/3}(t_1-t_2)^{-2}\, ,
\eea
where  $\alpha_{5,+} \equiv \alpha_{1,5} = -2\alpha_-$, $\alpha_{7,-}\equiv \alpha_{-1,-7} = \alpha_+ + 4\alpha_-$, $\alpha_+ \equiv \sqrt{10/3}, \alpha_- \equiv -\sqrt{3/10}$. 
There are four independent combination of contours, obtained by picking two (possibly identical) contours $\mathcal{C}_{1}$ and $\mathcal{C}_2$ out  of $ [0,z]$ and  $ [1,\infty)$. 
Out of these four combinations, only two are physical conformal blocks, corresponding to the exchange of the primary $\phi_{1,3}$ or $\phi_{1,5}$  allowed  by the fusion rules. We can identify the appropriate conformal block by noting that, in the $z \to 0$ limit, its leading power should be $z^{h_{1,i}-2h_{1,7}}$ with $i=3,5$.~\footnote{When  the integration interval is $[0,z]$, we can perform a   change of variables $t_i \to t_i/z$ in order to factor out `z'.} We identify our two conformal blocks: 
\bea
\label{eq:confblock-phi15}
N_{1,5}\mathcal{F}_{1,5}(z) & = z^{-7/5}(1-z)^{4/15}  \int_1^\infty dt_1\int_0^1dt_2 \frac{t_1^{2/5}(t_1-1)^{2/5}(t_1-z)^{2/5}}{t_2^{4/3}(t_2-1)^{4/3}(zt_2-1)^{4/3}(t_1-z t_2)^2}\, ,
 \\
\label{eq:confblock-phi13}
N_{1,3}\mathcal{F}_{1,3}(z) & = z^{-8/5}(1-z)^{4/15} \int_0^1 dt_1 dt_2 \frac{t_1^{2/5}(t_1-1)^{2/5}(z t_1-1)^{2/5}}{t_2^{4/3}(t_2-1)^{4/3}(zt_2-1)^{4/3}(t_1-t_2)^2}\, , 
\eea
where we  have added an arbitrary constant in the definition of the conformal blocks.
This coefficient can be fixed by requiring that  the first term of the integrands' series expansion in $z$ around zero is equal to one. 

Note however that the integrals  are not convergent because there are several not integrable singularities, 
such as   the point $t_2=1$ among many others.  
We then define the conformal blocks by    analytically continuing  the exponent `$4/3$' and `$2$' appearing in the integrands. 
Employing  this procedure for \reef{eq:confblock-phi15}, and expanding arround $z=0$,  yields
$
N_{1,5} = -\frac{e^{i\pi/3}\Gamma(-1/3)^2\Gamma(-1/5)\Gamma(7/5)}{\Gamma(-2/3)\Gamma(6/5)}
$.
 We can then perform the integral over the  first few terms in the $z$ expansion and, after analytically continuing, we obtain 
\be \label{series-phi15block-fewterms}
\mathcal{F}_{1,5}(z) \sim z^{-7/5}\left[1 + \frac{3}{10}z + \frac{208}{75}z^2 + \mathcal{O}(z^3)\right].
\ee
In practice this way of computing is not very efficient to compute large orders of the expansion. 
In the next section we will derive an efficient way of doing so by deriving a recursive relation for the coefficients above  from   the solutions of the BPZ equation. 

For the   $\phi_{1,3}$ conformal block  \reef{eq:confblock-phi13} we encounter a problem, we do not know how to define \reef{eq:confblock-phi13}
by an analytical continuation. 
We believe the root of the problem is the presence of the space-independent $1/(t_1-t_2)^2$ singularity, which in this case is due to the collision of the two screening charges.
We do not know a straightforward solution to this problem, and we would appreciate if a knowledgeable reader e-mails us with the solution.

 In any case for the calculation of Effective Hamiltonian, we will only need the $z=0$ expansion of these four-point functions, 
 and not the entire non-perturbative four-point function.
 For our purposes it will then be enough to solve the BPZ equation in a perturbative expansion around $z=0$, see appendix~\ref{BPZapp}.

 \section{Series expansions of the four-point function from BPZ equations }
 \label{BPZapp}

Next we discuss the calculation of the four-point function $\langle \phi_{1,5} \phi_{1,7} \phi_{1,7} \phi_{1,7}\rangle$.
 The conformal family descending from the field $\phi_{1,7}$ has a null state at level seven $\mathcal{L}^{\text{null}}_{1,7}\ket{\phi_{1,7}}=0$, 
 where, 
\bea
\mathcal{L}^{\text{null}}_{1,7} &= L_{-7}+\alpha_1 L_{-6}L_{-1}+\alpha_2L_{-5}L_{-2}+\alpha_3L_{-5}L_{-1}^2+\alpha_4L_{-4}L_{-3} \nonumber 
\eea
\bea
&+\alpha_5L_{-4}L_{-2}L_{-1}+\alpha_6L_{-4}L_{-1}^3 +\alpha_7L_{-3}^2L_{-1}+\alpha_8L_{-3}L_{-2}^2 \nonumber \\
&+\alpha_9L_{-3}L_{-2}L_{-1}^2+\alpha_{10}L_{-3}L_{-1}^4+\alpha_{11}L_{-2}^3L_{-1}+\alpha_{12}L_{-2}^2L_{-1}^3+\alpha_{13}L_{-2}L_{-1}^5+\alpha_{14}L_{-1}^7 \, 
, \label{eq:nullstate}
\eea
 and the coefficients are given by 
 \be
 \{\alpha_1,\alpha_2,\dots,\alpha_{14}\}=
 \{-\frac{20}{3}, \, \frac{3}{2}, \, -\frac{5}{4}, \,  2, \, -\frac{85}{6}, \, \frac{125}{24}, \, -\frac{19}{12}, \, -\frac{27}{10}, \, \frac{17}{4}, \, -\frac{35}{72}, \, 9, \, -\frac{245}{24}, \, \frac{175}{72}, \, -\frac{125}{864}\} 	\, . \nonumber
 \ee
 There is also a level sixth null state, either of the two can be used to construct a BPZ~\cite{Belavin:1984vu} differential equation obeyed by the correlator. 
 As a crosscheck, we have carried out both calculations, which of course give the same answer. 
 Here will present the seventh order differential equation.
 Since we are interested in the small $z$ expansion of the correlator  $\langle \phi_{1,5}(\infty) \phi_{1,7}(1) \phi_{1,7}(z) \phi_{1,7}(0)\rangle$
 it is impractical to use the null states of the $\phi_{1,5}$ field. 
 
 We parametrize the chiral part of the four-point function as
 \be \label{4ptfc-general}
 \langle \phi_{1,7}(z_0)\phi_{1,7}(z_1)\phi_{1,7}(z_2)\phi_{1,5}(z_3)\rangle = \frac{(z_0-z_3)^{2/15}(z_1-z_3)^{2/15}(z_2-z_3)^{2/15}}{(z_0-z_1)^{2/3}(z_0-z_2)^{2/3}(z_1-z_2)^{2/3}}\tilde{G}(\eta) \, , 
 \ee
 where as usual $\eta=\frac{(z_0-z_1)(z_2-z_3)}{(z_0-z_3)(z_2-z_1)}$.
 
Any correlator involving the null state must vanish. Such a correlator can be obtained from \eqref{4ptfc-general} via the Ward identities  (i.e.  identifiying the action of  the $L_n$'s in terms of contour integrals of the stress tensor): 
\[ \mathcal{L}_{-n} = \sum_{i=1}^3\left[\frac{(n-1)h_i}{(z_i-z_0)^n}-\frac{1}{(z_i-z_0)^{n-1}}\partial_{z_i}\right].\]
 Converting the null state expression \eqref{eq:nullstate} into a differential operator, applying it to the RHS of \eqref{4ptfc-general}, diving by the prefactor function and then sending $z_1 \to 0, z_2 \to 1, z_3 \to \infty, \eta \to z_0$, we obtain a seventh order linear ODE for the function $\tilde{G}$:
 \be
  \sum_{i=0}^7 p_i(z) \tilde{G}^{(i)}(z)=0 	\, ,  \label{eq:ODE7thorder}
 \ee
 where  we have defined $\tilde{G}^{(i)}(z) \equiv \frac{d^i}{dz^i}G(z)$, $\tilde{G}^{(0)}(z) \equiv \tilde{G}(z)$, and the functions $p_i$ are given by
\bea
  p_0(z) &= \frac{77077(52+91z-699z^2+848z^3-580z^4+246z^5-14z^6+4z^7)}{36905625(z-1)^7z^7}, \nonumber\\
 p_1(z) &= \frac{20447063+42471642z-41881164z^2-1038044z^3+161742z^4+428736z^5-142912z^6}{4920750(z-1)^6z^6},  \nonumber\\
 p_2(z)&=\frac{1436555-667796z-5860132z^2+1387388z^3+3779050z^4-1511620z^5}{109350(z-1)^5z^5},  \nonumber\\
 p_3(z)&=\frac{132299-2529964z-607083z^2+6274094z^3-3137047z^4}{43740(z-1)^4z^4},  \nonumber\\
 p_4(z)&=\frac{57530+108249z-669927z^2+446618z^3}{5832(z-1)^3z^3}, \nonumber\\
 p_5(z)&=-\frac{35(107-998z+998z^2)}{1296(z-1)^2z^2}, \ p_6(z) = -\frac{2275(-1+2z)}{1296(z-1)z},\  p_7(z) = -\frac{125}{864}\, .  \nonumber
\eea

We are interested in solving for the coefficients $\alpha_n$'s of the  
series representation of the four-point function:\footnote{From now one we work with the functions $G$, which differ from $\tilde{G}$ by an appropriate combination of OPE coefficients $\sqrt{C^s_{77}C^s_{57}}$, $s=3,5$ for the $\phi_{1,3},\phi_{1,5}$ conformal block, respectively.}
\be
G(z) = z^{x}\sum_{n\geq 0}\alpha_n z^n \label{diffans}\,.
\ee

 Inserting \reef{diffans} into the differential equation  and equating the lowest power of $z$ to zero yields the so called \textit{indicial equation}, whose solutions determine the possible values of $x$. For \eqref{eq:ODE7thorder} the indicial equation is proportional to
  \be \label{eq:indicial}
(x + 14/15) (x + 11/15) (x + 8/15) (x - 1/15) (x - 22/15) (x - 
   52/15) (x - 91/15)a_0 = 0
 \ee
 Each of these solutions correspond to the exchange of a scalar Kac primary field, but five of them are either disallowed by the fusion rules or are outside the Kac table of the $M(3,10)$.
  The solutions $x_{5,3}=-11/15, -14/15$ correspond to the $\phi_{1,5}, \phi_{1,3}$ block respectively. 
  
  Note that for the $x=-14/15$ solution, corresponding to the $\phi_{1,3}$ conformal block, there are two other solutions of the indicial equations which are an integer above $-14/15$: $1/15=-14/15+1$ and $91/15=-14/15+7$, corresponding to the hypothetical exchange of $\phi_{1,7}$ or $\phi_{1,13}$, which have holomorphic dimension $3/5=h_{1,3}+1$, $33/5=h_{1,3}+7$ respectively. This fact implies  that when solving for the coefficients of the $\phi_{1,3}$ block expansion, three coefficients will be left undetermined: $a_0,a_1$ and $a_7$. The coefficient $a_0$ is simply normalization, and in our conventions is set to one, while the coefficients $a_1$ and $a_7$ have to be obtained by another method. We will compute them analytically from first principles using the knowledge of the OPE coefficients of $\phi_{1,3}$ descendants up to level seven in the $\phi_{1,7}\times \phi_{1,7}$ OPE.

 \subsubsection*{Coefficients of descendant operators in the $\phi_{1,7}\times \phi_{1,7}$ OPE}
\label{77ope}

We are interested in the  
$\phi_{1,3}$ block of the $\phi_{1,7}\times \phi_{1,7}$ OPE:
\bea \label{eq:phi17OPE}
 \phi_{1,7}(z) &\times \phi_{1,7}(0)  =  |z|^{-2h_{1,7}}(\mathbb{1} + \cdots) + |z|^{h_{1,3}-2h_{1,7}}C^3_{77}\bigg(\phi_{1,3}(0)+zf_{L_{-1}}(L_{-1}\phi_{1,3})(0)+ h.c.\\
 & +z \bar{z} f_{L_{-1}}^2 (L_{-1}\bar{L}_{-1}\phi_{1,3})(0)+ z^2\big[f_{L_{-1}^2}(L_{-1}^2\phi_{1,3})(0)+f_{L_{-2}}(L_{-2}\phi_{1,3})(0)\big] \nonumber \\
 &+ z^2\bar{z}^2f_{L_{-1}^2}^2(L_{-1}^2\bar{L}_{-1}^2\phi_{1,3})(0)+z^2\bar{z}^2 f_{L_{-1}^2}f_{L_{-2}}\big[(L_{-1}^2\bar{L}_{-2}\phi_{1,3})(0)+h.c.\big] \\
 &+ z^2\bar{z}^2f_{L_{-2}}^2(L_{-2}\bar{L}_{-2}\phi_{1,3})(0)+\mathcal{O}(z^3,z^2\bar{z},\bar{z}^2z,\bar{z}^3)\bigg)+ |z|^{h_{1,5}-2h_{1,7}}C^5_{77}(\phi_{1,5}(0)+ \cdots)\,.
 \eea
 where each time we write `h.c.' means that one should add the anti-holomorphic analog of the preceeding term,
 for instance $zf_{L_{-1}}(L_{-1}\phi_{1,3})(0)+ h.c.= zf_{L_{-1}}(L_{-1}\phi_{1,3})(0)+\bar{z}f_{L_{-1}}(\bar{L}_{-1}\phi_{1,3})(0)$.
 
 Here we will use local Ward identities derived in Ref.~\cite{Ribault:2014hia} (see Eq.~2.2.56) to solve for the coefficients $f_L$ level by level. For a given pair of positive integers $n\leq N$, we have (specialising to the case of the $\phi_{1,3}$ block):
 \be \label{eq:local-ward-OPE}
 \sum_{|L|=N-n}f_L(h_{1,3}+N-n+(n-1)h_{1,7})(L\phi_{1,3})(0)=\sum_{|L|=N}f_L (L_nL\phi_{1,3})(0)\,.
 \ee
 By equating coefficients of various descendant operators on the LHS and RHS of \eqref{eq:local-ward-OPE}, one can solve for the $f_L, |L|=N$ from the knowledge of the $f_L, |L|=N-n$. In practice, choosing $n=1,2$ is enough to fix all coefficients recursively, given the seed $f_1 =1$. Note that the sums in \eqref{eq:local-ward-OPE} run over a basis of Virasoro raising operators for a given level, in particular they don't include all partitions of the given level if there are null states at this level. In this case, the Virasoro modes that are omitted from the basis have to be rexpressed in terms of those which form part of the basis, using the null combinations. For example, at level three in the $\phi_{1,3}$ family there is a null combination $\ket{\text{null}_{1,3}}\equiv (L_{-3}+5L_{-2}L_{-1}-25/6L_{-1}^3)\ket{\phi_{1,3}}$. If we take $L_{-3}\ket{\phi_{1,3}}$ and $L_{-2}L_{-1}\ket{\phi_{1,3}}$ as a basis at this level, then on the RHS of \eqref{eq:local-ward-OPE} for $N=5,n=2$ for example, the state $L_{-1}^3\ket{\phi_{1,3}}$ has to be replaced by $6/25(L_{-3}+5L_{-2}L_{-1})\ket{\phi_{1,3}}$ whenever it appears (for example in the term $L_2 L_{-3}L_{-1}^2$) in order to obtain a consistent solution for the $f$'s. 
 
 Solving Eq.~\eqref{eq:local-ward-OPE} for $n=1,2$ recursively for $N=2,3,4,5,6$ and $7$, we obtain the following OPE coefficients: 
\bea
&f_{L_{-1}}=1/2,\quad f_{L_{-1}^2}=-1/2,\quad f_{L_{-2}}= 1/6,\quad f_{L_{-3}} 1/75, \quad f_{L_{-2}L_{-1}}=-4/15, \quad  f_{L_{-4}}=-\frac{749}{8550}, \nonumber \\
 & f_{L_{-3}L_{-1}}=\frac{4931}{17100}, \quad f_{L_{-2}^2}=41/76,\quad f_{L_{-2}L_{-1}^2}=-\frac{8693}{6840}, \quad  f_{L_{-5}}=-\frac{2039}{142500}, \quad  f_{L_{-4}L_{-1}}=\frac{5451}{19000}, \nonumber\\
&  f_{L_{-3}L_{-2}}=\frac{55931}{142500},\quad f_{L_{-3}L_{-1}^2}=-\frac{2119}{4750},\quad f_{L_{-2}^2L_{-1}}=-\frac{26513}{57000}, \quad f_{L_{-6}}=-\frac{31767}{688750}, \nonumber
\eea
\bea
 &  f_{L_{-5}L_{-1}}=\frac{209539}{1377500}, \quad f_{L_{-4}L_{-2}}=\frac{2208}{14463750},\quad f_{L_{-4}L_{-1}^2}=-\frac{984313}{7714000}, \quad  f_{L_{-3}^2}=\frac{17672873}{289275000}, \nonumber \\
&  f_{L_{-3}L_{-2}L_{-1}}=-\frac{4480117}{28927500},\quad f_{L_{-2}^3}=\frac{1138}{11571},\quad f_{L_{-2}^2L_{-1}^2}=-\frac{2441373}{7714000}, \quad  f_{L_{-7}}=-\frac{168347}{9642500}, \nonumber \\
 & f_{L_{-6}L_{-1}}=\frac{357424}{2410625},\quad f_{L_{-5}L_{-2}}=\frac{631079}{5165625},\quad f_{L_{-5}L_{-1}^2}=-\frac{63507}{507500},\quad  f_{L_{-4}L_{-3}}=\frac{952487}{28927500}, \nonumber \\
 & f_{L_{-4}L_{-2}L_{-1}}=\frac{349607}{28927500},\quad f_{L_{-3}^2L_{-1}}=-\frac{113801}{6887500},\quad f_{L_{-3}L_{-2}^2}=\frac{17173691}{144637500}, \nonumber \\
&f_{L_{-3}L_{-2}L_{-1}^2}=-\frac{2216421}{9642500},\quad f_{L_{-2}^3L_{-1}}=-\frac{2741309}{28927500}\, . \nonumber
\eea

 The coefficients of the $z=0$ expansion of the $\phi_{1,3}$ conformal block can then be obtained by summing over the $f$ coefficients at a given level, weighted by the three point function containing the corresponding descendant. For example, the third coefficient in the expansion, i.e. the one with power $z^{h_{1,3}-2h_{1,7}+2}$ is equal to (modulo the $C^3_{77}$ OPE coefficient which we usually factor outside the conformal block): $f_{L_{-1}^2}\langle\phi_{1,5}(\infty)\phi_{1,7}(1)(L_{-1}^2\phi_{1,3})(0)\rangle  +f_{L_{-2}}\langle\phi_{1,5}(\infty)\phi_{1,7}(1)(L_{-2}\phi_{1,3})(0)\rangle$. In this way, we obtain from first principles the first few coefficients of the $\phi_{1,3}$ conformal block: 
 \begin{align}
 \begin{aligned} \label{eq:phi13-confblock} 
 \mathcal{F}_{1,3}(z)=z^{-8/5}\left[1+\frac{1}{5}z-\frac{17}{150}z^2-\frac{24}{125}z^3-\frac{29179}{71250}z^4-\frac{192269}{296875}z^5\right.\\
 \left.-\frac{38223871}{43046875}z^6-\frac{242271808}{215234375}z^7+\mathcal{O}(z^8)\right]\,.
 \end{aligned}
 \end{align}
 On the other hand, the conformal block can also be computed from the series solution of the BPZ:
 \be \label{eq:F13-series}
\mathcal{F}_{1,3}(z) = \frac{G_3(z)}{z^{2/3}(1-z)^{2/3}}=\frac{z^{x3}\sum_{n=0}^\infty a^3_n z^n}{z^{2/3}(1-z)^{2/3}}.
  \ee
 By comparing \eqref{eq:phi13-confblock} to \eqref{eq:F13-series}, one can fix the coefficients $a^3_1$ and $a^3_7$.,\footnote{The other coefficients $a^3_i, i\leq6, i\neq 0,1$ are fixed by the recursion relation once $a_0^3$ and $a_1^3$ are fixed, and are found to match with \eqref{eq:phi13-confblock}, which is a non trivial crosscheck.} and from this use the BPZ equation to recursively compute $a^3_n$'s to high order. For the $\phi_{1,5}$ block, once normalization is imposed the $a^5_n$'s can likewise be extracted from the BPZ recursively to high order. All in all, we obtain the $a_m^s$ ($s=3,5$) in the following expansion of the correlator:
 \be 
 \langle \phi_{1,5}(\infty)\phi_{1,7}(1)\phi_{1,7}(z,\bar{z})\phi_{1,7}(0)\rangle =\sum_{s=3,5}C^s_{57}C^s_{77}|\mathcal{F}_s(z)|^2=\sum_{s=3,5}C^s_{57}C^s_{77} \Bigg|\frac{z^{x_s}\sum\limits_{m=0}^\infty a^s_m z^m }{z^{2/3}(1-z)^{2/3}}\Bigg|^2\,.
 \ee
 The $x_s$ are the relevant solutions to the indicial equation \eqref{eq:indicial}.
We now describe how to determine the $B_{K,Q}$ coefficients defined in \eqref{eq:def-BKQ}. Indeed, for the computation of $K_{\text{eff},3}$, we need the four-point function evaluated at variable positions $\langle \phi_{1,5}(\infty)\phi_{1,7}(1)\phi_{1,7}(z_1,\bar{z}_1)\phi_{1,7}(z_2,\bar{z}_2)\rangle$. This correlator can be expanded as:
\be 
 \sum_{s=\{1,3\},\{1,5\}}C^s_{57}C^s_{77}
\sum_{n_1,n_2,n_3,m=0}^\infty a^s_m r_{n_1}^{2/15+h_s+m}r_{n_2}^{2/3}r_{n_3}^{6/5-h_s-m}\left(\frac{z_1}{z_2}\right)^{n_1+n_3}z_2^{n_1+n_2+m-6/5+h_s}\times \text{c.c.} \, .
\ee
For fixed powers of $z_1/z_2, z_2$ (which we label respectively $K,Q$) we can first fix $m \leq Q$ and perform the sum over $n_1$, defining
\begin{align*} 
A_{M,N}^s(m)\equiv \sum_{i=0}^{\text{min}(M,N)}r_i^{2/15+h_s+m}r_{M-i}^{6/5-h_s-m}r_{N-i}^{2/3}= \frac{\Gamma(6/5-h_s-m+M)\Gamma(2/3+N)}{\Gamma(2/3)\Gamma(6/5-h_s-m)\Gamma(1+M)\Gamma(1+N)}\\
_3F_2(2/15+h_s+m,-M,-N;-1/5+h_s+m-M,1/3-N;1).
\end{align*}
Then, the $B_{K,Q}^s$ coefficients are obtained by appropriately summing the $A_{K,Q-m}(m)$ over $m$:
 \be \label{eq:5777-BKQ}
 B^s_{K,Q} \equiv \sum_{m=0}^Qa_m^sA^s_{K,Q-m}(m)\,. 
 \ee

\subsection*{$\langle \phi_{1,3} \phi_{1,7} \phi_{1,7} \phi_{1,7}\rangle$}

As explained in Sec. \ref{CGapp}, the $\langle \phi_{1,3} \phi_{1,7} \phi_{1,7} \phi_{1,7}\rangle$ correlator can be computed via the Coulomb gas formalism, and is given in Eq. \eqref{eq:4ptfc-phi13}. 
The coefficients $B_{K,Q}$ can directly be extracted from the closed form expression and are equal to:
\begin{align}
\begin{aligned} \label{eq:3777-BKQ}
 B_{K,Q} \equiv &A_{K,Q}-A_{K-1,Q-1}-A_{K,Q-1}+A_{K-2,Q-2}-A_{K-1,Q_2}+A_{K,Q-2},\\
 A_{K,Q} \equiv &\sum_{k=0}^{\text{Min}(K,Q)} r_{K-k}^{7/5}r_{Q-k}^{7/5}r_k^{7/5}, \quad (A_{K,Q} = 0, \, K \text{ or } Q < 0)\,.
\end{aligned}
\end{align}
The $A_{K,Q}$'s admit the closed form representation:
\be
 A_{K,Q}= \frac{\Gamma(K+7/5)\Gamma(Q+7/5)\, _3F_2(7/5,-K,-Q;-K-2/5,-Q-2/5;1)}{(\Gamma(7/5)^2\Gamma(K+1)\Gamma(Q+1)}.
\ee
One can of course also compute these coefficients via a BPZ ODE, which we have done as a crosscheck. We find agreement with \eqref{eq:3777-BKQ}.

\section{More details regarding the contributions to $K_{\text{eff}}$ from $\phi_{1,3}$ descendants}
\label{desct}

In this section we provide more detail regarding the contributions to $K_{\text{eff}}$ from the leading descendants of $\phi_{1,3}$. We are interested in the following terms in the $\phi_7 \phi_7$ OPE:
\be \label{eq:OPE-desc}
\phi_{1,7}(z,\bar{z})\phi_{1,7}(w,\bar{w}) \supset \frac{C^3_{77}}{|z-w|^{16/5}}\left[\frac{1}{2}(z-w)(L_{-1}\phi_{1,3})(w,\bar{w})+\frac{1}{2}(\bar{z}-\bar{w})(\bar{L}_{-1}\phi_{1,3})(w,\bar{w})\right]\, .
\ee
As explained in Sec. \ref{ptrenorm}, these descendants in the OPE are associated with UV divergences at second order in $g_7$, and therefore require counterterms. While the UV divergences are best analyzed via singularities of plane correlation functions, counterterms are added to the Hamiltonian defined on the cylinder, we therefore need to know the Weyl transformation properties of the level one descendants. By following an analogous derivation to that of the identity descendants $T,\bar{T}$ (see for example \cite{DiFrancesco:1997nk}), it is possible to show that the holomorphic level one descendant of $\phi_{1,3}$ on the plane is related to the same operator on the cylinder as:
\be \label{eq:transf-desc}
(L_{-1}\phi_{1,3,\text{cyl.}})(w,\bar{w})=R^{-1/5}|z|^{-4/5}\Big[z (L_{-1}\phi_{1,3,\text{pl.}})(z,\bar{z})-\frac{2}{5}\phi_{1,3,\text{pl.}}(z,\bar{z})\Big].
\ee
Here subscripts indicate on which manifold an operator is defined, and we use cylinder complex coordinates defined in terms of the usual Weyl map from plane coordinates $w=R\ln{z}= \tau + i x$. From Eqs. \eqref{entwo}, \eqref{eq:OPE-desc} and \eqref{eq:transf-desc}, one can derive the appropriate counterterm to add to the bare theory \eqref{eq:QFT-Hamiltonian} to cancel this particular divergence:
\begin{align}
\label{eq:ct-desc}
V_{\text{c.t.}}^{(2)}(\epsilon)\supset R^{2-2\Delta_7+\Delta_3+1} \left(\frac{g_7}{2\pi}\right)^2C^3_{77}\int_{\substack{|z| \leq 1 \\ |1-z|>\eps}} d^2z |z|^{\Delta_7-2} \frac{z-1}{|1-z|^{16/5}}\int_0^{2\pi R}dx \mathcal{O}_{\text{cyl.}}(0,x)\,,\\
 \mathcal{O}_{\text{cyl.}}(\tau,x) = \frac{1}{2}\left[L_{-1}\phi_{1,3,\text{cyl.}}(\tau,x)+\bar{L}_{-1}\phi_{1,3,\text{cyl.}}(\tau,x)\right]+\frac{2}{5}R^{-1}\phi_{1,3,\text{cyl.}}(\tau,x)
 \,.
\end{align} 
Now, the cylinder operator $L_{-1}\phi_{1,3,\text{cyl.}}(\tau,x)+\bar{L}_{-1}\phi_{1,3,\text{cyl.}}(\tau,x)$ has vanishing matrix elements between scalar states. The remaining piece of $\mathcal{O}_{\text{cyl.}}$ is proportional to the $\phi_{1,3}$ primary, hence the claim in Sec. \ref{ptrenorm} that the UV divergence proportional to the level one descendants of $\phi_{1,3}$ can be interpreted as a subleading divergence in $\phi_{1,3}$. Note the correspondance with Eq. \eqref{ct2des} upon identifying $\langle \varphi(\infty)\phi_7(1)\phi_7(z,\bar{z})\rangle= C^3_{77}(z-1)/(2|1-z|^{16/5})$.

Let us now move on the Effective Hamiltonian. The counterterm in \eqref{eq:ct-desc} has been exactly crafted to be equal and opposite the sum over high energy states from the local contribution in $H_{\text{eff},2}(\epsilon)$ proportional to the descendant. As a consequence:
\begin{align} \label{Keff2-desc}
\left(H_{\text{eff},2}(\epsilon) + \eqref{eq:ct-desc}\right)_{fi}\supset \frac{g_7^2R^{3-2\Delta_7}}{2} & \Bigg[\int_{\substack{|z| \leq 1 \\ |1-z|>\eps}} d^2z |z|^{\Delta_7-2} \langle \varphi(\infty)\phi_7(1)\myline_i\phi_7(z,\bar{z})\rangle  \\ & + \int_{\substack{|z| \leq 1 \\ |1-z|>\eps}} d^2z |z|^{\Delta_7-2} \langle \bar{\varphi}(\infty)\phi_7(1)\myline_i\phi_7(z,\bar{z})\rangle  \Bigg](\mathcal{O}_{\text{cyl.}})_{fi}+f\leftrightarrow i\,.\nonumber 
\end{align}
We have only written out the contribution from the descendants, which now explicitly depend only on the low energy states, since $\myline_i = \sum_{\Delta_l\leq \Delta^T_{7,i}}\ket{l}\bra{l}$. Expanding \eqref{Keff2-desc} in powers of $z$ and $\bar{z}$, integrating and taking the $\epsilon \to 0$ limit we precisely recover the $\beta^3_{\Delta^T_{7,i}}(\phi_3)_{fi}$ term in Eq. \eqref{keff2}.

\section{Details of EFT computations}
\label{EFT-computations}

\subsection*{Matrix elements of $L_{-2}\bar{L}_{-2}\phi$}\label{eq:2ndlevel-matel}

This appendix includes the matrix element computation of various operators we add to the EFTs we use to fit in the main text, i.e. $L_{-2}\bar{L}_{-2}\phi$ and $T\bar{T} = L_{-2}\bar{L}_{-2}\mathbb{1}$.
This section generalises and adapts to our needs the EFT computations presented in \cite{Xu:2022mmw}.

 We are interested in computing the matrix elements of the scalar descendant operator $L_{-2}\bar{L}_{-2}\phi_{h,h}$ in a generic Minimal Model, between scalar states:
\be \bra{i}\int_0^{2\pi R} dxL_{-2}\bar{L}_{-2}\phi_{h,h}(0,x)\ket{i}.\ee
The descendant operators on the cylinder are defined as (the $\bar{L}_n$ descendants are defined in an analogous way):
\be 
L_n \mathcal{O}(z_0,\bar{z}_0)\equiv \oint_{\mathcal{C}_{z_0}}\frac{dz}{2\pi i}T(z)(z-z_0)^{n+1}\mathcal{O}(z_0,\bar{z}_0)\,,
\ee
where $z=x+iy$ is a complex coordinate over the cylinder, $x \sim x+2\pi R, y \in \mathbb{R}$. Since we are computing the matrix element of a scalar operator between scalar states, we can simply evaluate the descendant at the origin of the cylinder, and the $x$ integral over space gives a trivial overall volume factor $2\pi R$. 

We will compute the contour integral of the stress tensor by replacing $(z)^{-1}$ by $U_2(z)$, a $2\pi$-periodic function of $z$ which has the same Laurent series up till terms of order $z^2$ or higher.~\footnote{These terms correspond to operators of the form $L_{n\geq 1}\mathcal{O}$ which vanish for $\mathcal{O}$ primary.} We will choose this function to be: 
\[ U_2(z) \equiv \frac{1}{2}\frac{\cos{(z/2)}}{\sin{(z/2)}}+\frac{\sin{z}}{12} = \frac{1}{z}+ \mathcal{O}(z^3)\,.\]
This will allow us to rewrite the expression into a sum of matrix elements involving the primary operator $\phi_{h,h}$ and the Virasoro generators $\bm{L}_n,\bar{\bm{L}}_n$ which act on the cylinder Hilbert space, and are defined as:
\be \bm{L}_{n} \equiv -R \int_0^{2\pi R}\frac{dz}{2\pi}T(z)e^{-\tfrac{inz}{R}}+\frac{c}{24}\delta_{n,0},
\ee
where $c$ is the central charge. 
Now, since the matrix element factorizes into a holomorphic and antiholomorphic component, let us focus on the holomorphic part:
\[ \bra{i} \int_{\mathcal{C}_{0}}\frac{dz}{2\pi i}\frac{U_2(z/R)}{R} \mathcal{R}\{T(z)\phi_{h,h}(0)\}\ket{i}\,,\]
where we have explicitly written the implicit radial ordering whenever more than one cylinder operator is present in a correlation function. The contour integral around $0$ of the radially ordered expression can be written as:
\[\underbrace{\bra{i} \int_{0,\text{Im}(z)<0}^{2\pi R}\frac{dx}{2\pi i} \frac{U_2(z/R)}{R}T(z)\phi_{h,h}(0)\ket{i}}_{A_i} - \underbrace{\bra{i}\int_{0,\text{Im}(z)>0}^{2\pi R}\frac{dx}{2\pi i} \frac{U_2(z/R)}{R}\phi_{h,h}(0)T(z)\ket{i}}_{B_i}\,.\]
Now, we write the Fourier expansion of the function $U_2(z)$ in the two cases $y \lessgtr 0$:
\[ U_2(z)-\frac{\sin{z}}{12} = \sum_{n\geq0} c_n^\lessgtr e^{i n z}, \quad \text{Im}(z)\lessgtr 0, \quad c_{n>0}^>=-i, \, c_0^>=-i/2, \quad c_n^< =- c_{-n}^>\,. \]
We have ommitted the trivial fourier expansion of the sine in the definition of the $c_n$'s, which one can add back by hand in the formulas. Now, plugging in the Fourier decomposition of $U_2(z/R)/R$ in the expression for the $A_i$'s and $B_i$'s, we obtain:
\be
 A_i = \bra{i} \int_0^{2\pi R} \frac{dx}{2\pi R i}\left[\sum_{n\leq 0} c_n^< e^{inz/R} +\frac{1}{12}\frac{1}{2i}(e^{iz/R}-e^{-iz/R})\right]T(z) \phi_{h,h}(0,0)\ket{i}
\ee
Similarly, the $B_i$'s read:
\be
 B_i = \bra{i} \int_0^{2\pi R} \frac{dx}{2\pi R i} \phi_{h,h}(0,0)\left[\sum_{n\geq 0} c_n^> e^{inz/R} +\frac{1}{12}\frac{1}{2i}(e^{iz/R}-e^{-iz/R})\right]T(z)\ket{i} \, .
\ee
Next, using $[\bm{L}_n,\phi_{h,h}(0,0)]=n h \phi_{h,h}(0,0)+[\bm{L}_0,\phi_{h,h}(0,0)]$, we can write:
\begin{align}
\begin{aligned}
A_i-B_i = \frac{1}{R^2}\Big[\bra{i}\left(-\left(\bm{L}_0-\frac{c}{24}\right)+\frac{h}{12}\right)\phi_{h,h}(0,0)\ket{i}\\
-\sum_{n>0}\bra{i}\bm{L}_n\phi_{h,h}(0,0)\ket{i}-\sum_{n>0}\bra{i}\phi_{h,h}(0,0)\bm{L}_{-n}\ket{i}\Big] \,.
\end{aligned}
\end{align}

To evaluate the previous expression, let's start by assuming $\ket{i} = \ket{\phi_i}$ a primary state with holomorphic dimension $h_i$. In this case, $\bm{L}_{n>0}$ annihilates $\ket{i}$ (likewise $\bm{L}_{n<0}$ annihilates $\bra{i}$), therefore we can replace $\bm{L}_n\phi_{h,h}(0,0)$ (respectively $\phi_{h,h}(0,0)\bm{L}_{-n}$) by the commutator, and we obtain:
\be \bra{i}\int_0^{2\pi R} dx L_{-2}\bar{L}_{-2}\phi_{h,h}(0,x)\ket{i}=2\pi C^h_{ii} R^{-2h-3}
\left[\left(-\left(h_i-\frac{c}{24}\right)+\frac{h}{12}\right)+\frac{h}{6}\right]^2,
\ee
with $C^3_{ii}$ the OPE coefficient between the primary $\phi_{h,h}$ and two copies of $\phi_i$, and we have used zeta function regularization $\sum_{n>0}n = \zeta(-1) = -1/12$. In the last step  we added back the anti-holomorphic contribution, which contributes the same as the holomorphic part for a scalar primary external state.

Next let' us consider the case where $\ket{i}= \bm{L}_{-1}\bar{\bm{L}}_{-1}\ket{\phi_i}$. Now the $n=1$ term in the sums still give a non zero contribution after commutation, which reads: 
\be
\sum_{n\geq1}\bra{\phi_i}\bm{L}_1 \bm{L}_n\phi_{h,h}(0,0)\bm{L}_{-1}\ket{\phi_i}=\left[-\tfrac{h}{12}(h(h-1)+2h_i)+2h h_i\right]\bra{\phi_i}\phi_{h,h}(0,0)\ket{\phi_i}.
\ee
Putting everything together:
\begin{align}
 \bra{i}\int_0^{2\pi R} dx L_{-2}\bar{L}_{-2}\phi_{h,h}(0,x)\ket{i}=2\pi C^h_{ii} R^{-2h-3}
\Big(\big[-\left(h_i+1-\frac{c}{24}\right)+\frac{h}{12}\big]\left(h(h-1)+2h_i\right)\\
-2\left(-\frac{h}{12}(h(h-1)+2h_i)+2h h_i\right)\Big)^2,
\end{align}
However, the descendant state $\bm{L}_{-1}\bar{\bm{L}}_{-1}\ket{\phi_i}$ is not normalized, having norm squared $4h_i^2$. Therefore, when computing first order perturbation theory for EFT fits, we must divide the above equation by the norm squared of the state. Putting the $M(3,8)$ values of $c=-21/4, h=h_{1,3}=-1/4$ and $h_i=\{h_{1,3},h_{1,5}\}=\{-1/4,1/4\}$, and after factoring by the appropriate OPE coefficients and volume factors, we readily obtain the values quoted in \eqref{EFT-coeffs-m38}.

Lastly, note that by setting $h=0$ in the previous formulas one recovers the fact that the first order correction from $T\bar{T}=L_{-2}\bar{L}_{-2}\mathbb{1}$ to a state $\ket{\Delta_i}$ is universal and proportional to $(\Delta_i/2-\frac{c}{24})^2$. 

\subsection*{Corrections to gaps in $M(3,5)$ and $M(3,7)$ EFT}
 
 In this section we provide more details on the EFT fits based on the actions \eqref{action-M35} and \eqref{action-M37}. We start by giving the CFT data of the even sector of both CFTs. The spectrum of primaries is shown in Tabs. \ref{tab:M35-evenops} and \ref{tab:M37-evenops}. The fusion rules in the even sector of the $M(3,5)$ simply amount to:
 \be \label{fusion-M35}
 \phi_{1,3}\times \phi_{1,3}\sim \mathbb{1}\,,
 \ee
 while those for the $M(3,7)$ are:
 \be \label{fusion-M37}
 \phi_{1,3}\times \phi_{1,3}\sim \mathbb{1} + i\phi_{1,3}+\phi_{1,5}, \quad \phi_{1,3}\times \phi_{1,5}\sim \phi_{1,3}+i\phi_{1,5},\quad \phi_{1,5}\times \phi_{1,5}\sim \mathbb{1}+i\phi_{1,3}\,.
 \ee
 Finally, the non trivial OPE coefficients in the even sector of the $M(3,7)$ are:\footnote{See \cite{Delouche:2023wsl} for exact values.}
 \be \label{OPE-M37}
 C^3_{33} \approx 1.87i,\quad C^3_{55}\approx 0.98i,\quad C^3_{35}\approx -0.11\,.
 \ee
 
\begin{table}[h]
    \centering
    \begin{tabular}{l l l l}
 $M(3,5)$ $\mathbb{Z}_2$-even ops.       \phantom{--} & $\phi_{1,1}$   \phantom{--} & $\phi_{1,3}$ \phantom{\Big|} \\
\hline
\hline
    $\Delta$ &  0  & 2/5 \phantom{\Big|}\\
\hline
$\mathcal{PT}$ & Even & Odd \phantom{\Big|} \\
\hline
    \end{tabular}
    \caption{The  $\mathbb{Z}_2$-even scalar primaries of the $M(3,5)$  model, their dimensions and ${\cal PT}$-symmetry.}
    \label{tab:M35-evenops}
\end{table}

\begin{table}[h]
    \centering
    \begin{tabular}{l l l l}
 $M(3,7)$ $\mathbb{Z}_2$-even ops.       \phantom{--} & $\phi_{1,1}$   \phantom{--} & $\phi_{1,3}$ \phantom{--} & $\phi_{1,5}$ \phantom{\Big|} \\
\hline
\hline
    $\Delta$ &  0  & -2/7 & 8/7\phantom{\Big|}\\
\hline
$\mathcal{PT}$ & Even & Odd  & Even \phantom{\Big|} \\
\hline
    \end{tabular}
    \caption{The  $\mathbb{Z}_2$-even scalar primaries of the $M(3,7)$  model, their dimensions and ${\cal PT}$-symmetry.}
    \label{tab:M37-evenops}
\end{table}
 
 \subsubsection*{$M(3,5)$ EFT}
 
 Based on the spectrum of the $M(3,5)$ CFT, the lowest four gaps in the even sector correspond to the states $\{\ket{\phi_{1,3}}, \ket{L_{-1}\bar{L}_{-1}\phi_{1,3}}, \ket{T\bar{T}}, \ket{L_{-2}\bar{L}_{-2}\phi_{1,3}}\}$\footnote{The fourth gap in the CFT is part of a four dimensional degenerate subspace generated by $\{\ket{L_{-1}^2\bar{L}_{-1}^2\phi_{1,3}},\ket{L_{-1}^2\bar{L}_{-2}\phi_{1,3}},\ket{L_{-2}\bar{L}_{-1}^2\phi_{1,3}},\ket{L_{-2}\bar{L}_{-2}\phi_{1,3}}\}$. Note that this degeneracy is not lifted by the $T\bar{T}$ deformation at tree level, but only by the $\phi_{1,3}$ deformation. To compute the first order correction to the fourth gap from $\phi_{1,3}$, we use degenerate perturbation theory in the degenerate subspace, and pick the state receiving the most negative correction, which we choose to label $\ket{L_{-2}\bar{L}_{-2}\phi_{1,3}}$.}, with scaling dimensions $\{2/5,2+2/5,4,4+2/5\}$. The fusion rule \eqref{fusion-M35} imply that the EFT operator $\phi_{1,3}$ doesn't correct any of the states at tree level, we therefore include its second order contribution. For the first three graps, we use non degenerate perturbation theory to numerically compute the correction \cite{Delouche:2023wsl}. For the fourth gap, we do the same using degenerate perturbation theory. All in all, combining with the $T\bar{T}$ corrections computed in the previous section, we obtain:

 \begin{align}
RG_{\phi_{1,3}}=&\Delta_3+\frac{g_{T\bar{T}}}{4(MR)^2}\left[\left(\Delta_3-\frac{c}{12}\right)^2-\left(\frac{c}{12}\right)^2\right]+(i g_3)^2(MR)^{8/5}(\gamma_{\phi_{1,3}}-\gamma_{\mathbb{1}})\,,\label{eq:fitM35-1}\\
RG_{\partial\bar{\partial}\phi_{1,3}}=&\Delta_3+2+\frac{g_{T\bar{T}}}{4(MR)^2}\left[\left(\Delta_3+2-\frac{c}{12}\right)^2-\left(\frac{c}{12}\right)^2\right]+\frac{(i g_3)^2}{(MR)^{-8/5}}(\gamma_{\partial\bar{\partial}\phi_{1,3}}-\gamma_{\mathbb{1}})\,,\\
RG_{T\bar{T}}=&4+\frac{g_{T\bar{T}}}{4(MR)^2}\left[\left(4-\frac{c}{12}\right)^2-\left(\frac{c}{12}\right)^2\right]+(i g_3)^2(MR)^{8/5}(\gamma_{T\bar{T}}-\gamma_{\mathbb{1}})\,,\\
RG_{\partial\partial\bar{\partial}\bar{\partial}\phi_{1,3}}=&\Delta_3+4+\frac{g_{T\bar{T}}}{4(MR)^2}\left[\left(\Delta_3+4-\frac{c}{12}\right)^2-\left(\frac{c}{12}\right)^2\right]+\frac{(i g_3)^2}{(MR)^{-8/5}}(\gamma_{\partial\partial\bar{\partial}\bar{\partial}\phi_{1,3}}-\gamma_{\mathbb{1}})\,.\label{eq:fitM35-2}
 \end{align}
 The $\gamma$ coefficients are given as:
 \be \label{gamma-coeffs}
\gamma_{\mathbb{1}}\approx -2.6,\,\gamma_{\phi_{1,3}}\approx 2.48,\, \gamma_{\partial\bar{\partial}\phi_{1,3}}\approx -0.41\,,\gamma_{T\bar{T}}\approx -15.3,\gamma_{\partial\partial\bar{\partial}\bar{\partial}\phi_{1,3}}\approx 1.70\,.
 \ee
 The fit shown in the left pannel of Fig. \ref{fits2} is performed by setting $M=1/R$ and minimising the chi-squared function $\chi^2_{M(3,5)}(g_{T\bar{T}},g_3)=R^2\sum_i \left(G_i(g_{T\bar{T}},g_3)-\Delta E_i\right)^2$, where the index $i$ runs over the four gaps \eqref{eq:fitM35-1}-\eqref{eq:fitM35-2}. The $\Delta E_i$'s are the HT gaps assuming the identity is the lowest energy state, as is the case in the even sector of the $M(3,5)$.
 
 \subsubsection*{$M(3,7)$ EFT}
 
 Based on the spectrum of the $M(3,5)$ CFT, the lowest four gaps in the even sector correspond to the states $\{\ket{\phi_{1,3}}, \ket{\phi_{1,5}},\ket{L_{-1}\bar{L}_{-1}\phi_{1,3}}, \ket{L_{-1}\bar{L}_{-1}\phi_{1,5}}\}$ of scaling dimensions $\{-2/7,8/7,2-2/7,2+8/7\}$ respectively. In the EFT \eqref{action-M37}, we have chosen to include only the leading two irrelevant $\bZ_2$-even irrelevant operators: $L_{-2}\bar{L}_{-2}\phi_{1,3}$ and $L_{-2}\bar{L}_{-2}\mathbb{1}=T\bar{T}$. Using the results of sec. \ref{eq:2ndlevel-matel}, the corrections to the first four gaps read:
 
 \begin{align}
 RG_{\phi_{1,3}}=&\Delta_3+\frac{g_{T\bar{T}}}{4(MR)^2}\left[\left(\Delta_3-\frac{c}{12}\right)^2-\left(\frac{c}{12}\right)^2 \right]+\frac{i g_3'}{(MR)^{12/7}}C^3_{33}\delta_{\phi_{1,3}}\,,\label{eq:fitM37-1} \\
 RG_{\phi_{1,5}}=&\Delta_5+\frac{g_{T\bar{T}}}{4(MR)^2}\left[\left(\Delta_5-\frac{c}{12}\right)^2-\left(\frac{c}{12}\right)^2 \right]+\frac{i g_3'}{(MR)^{12/7}}C^3_{55}\delta_{\phi_{1,5}}\,, \\
 RG_{\partial\bar{\partial}\phi_{1,3}}=&\Delta_3+2+\frac{g_{T\bar{T}}}{4(MR)^2}\left[\left(\Delta_3+2-\frac{c}{12}\right)^2-\left(\frac{c}{12}\right)^2 \right]+\frac{i g_3'}{(MR)^{12/7}}C^3_{33}\delta_{\partial\bar{\partial}\phi_{1,3}}\,, \\
  RG_{\partial\bar{\partial}\phi_{1,5}}=&\Delta_5+2+\frac{g_{T\bar{T}}}{4(MR)^2}\left[\left(\Delta_5+2-\frac{c}{12}\right)^2-\left(\frac{c}{12}\right)^2 \right]+\frac{i g_3'}{(MR)^{12/7}}C^3_{55}\delta_{\partial\bar{\partial}\phi_{1,5}}\,.\label{eq:fitM37-2}
 \end{align}
 The $\delta$ coefficients are:
 \be \label{delta-coefs}
 \delta_{\phi_{1,3}}=\frac{1}{576}\approx 0.0017,\,\, \delta_{\phi_{1,5}}=\frac{16129}{28224}\approx 0.57,\,\, \delta_{\partial\bar{\partial}\phi_{1,3}}=\frac{81}{3136}\approx 0.026,\,\, \delta_{\partial\bar{\partial}\phi_{1,5}}=\frac{64009}{21609}\approx 2.96\,.
 \ee
  The fit shown in the right pannel of Fig. \ref{fits2} is performed by setting $M=1/R$ and minimising the chi-squared function $\chi^2_{M(3,7)}(g_{T\bar{T}},g_3)=R^2\sum_i \left(G_i(g_{T\bar{T}},g_3)-\Delta E_i\right)^2$, where the index $i$ runs over the four gaps \eqref{eq:fitM37-1}-\eqref{eq:fitM37-2}. The $\Delta E_i$'s are the HT gaps assuming the identity is the second lowest energy state, as is the case in the even sector of the $M(3,7)$.
 
 \newpage 
\bibliography{biblio}
\bibliographystyle{utphys}

\end{document}